\documentclass[useAMS,usenatbib,usegraphicx]{mn2e}
\usepackage{times}
\usepackage{natbib}
\usepackage{amsmath}
\usepackage[dvips]{graphicx}
\usepackage{color}
\usepackage{subfigure}
\usepackage{multirow}

\newcommand{\divb}{ {\bf \nabla}\cdot{\bf B}}
\newcommand{\Bf}{{\bf B}}

\def\divB{\divb}

\newcommand{\kpc}{\mathrm{kpc}}

\def\bee{\begin{equation}}
\def\eee#1{\label{eq:#1}\end{equation}}
\def\bea{\begin{eqnarray}}
\def\eea#1{\label{eq:#1}\end{eqnarray}}

\def\EQ#1{Eq. (\ref{eq:#1})}

\newcommand{\FIG}[1]{Fig. (\ref{fig:#1})}
\newcommand{\FIGp}[1]{Fig. \ref{fig:#1}}
\newcommand{\bFIG}{\begin{figure*}}
\newcommand{\eFIG}[1]{\label{fig:#1}\end{figure*}}
\newcommand{\bFIGs}{\begin{figure}}
\newcommand{\eFIGs}[1]{\label{fig:#1}\end{figure}}

\title[Divergence-cleaning and SPMHD]{A divergence-cleaning scheme for cosmological SPMHD simulations}
\author[F. A. Stasyszyn et al.]{F. A. Stasyszyn$^{1}$\thanks{E-mail: fstasys@usm.lmu.de}, K. Dolag$^{1,2}$ and A. M. Beck$^{1,3}$\\
  $^{1}$University Observatory Munich, Scheinerstr. 1, D-81679 Munich, Germany\\
  $^{2}$Max Planck Institute for Astrophysics, Karl-Schwarzschild-Str. 1, D-85741 Garching, Germany\\
  $^{3}$Max Planck Institute for Extraterrestrial Physics, Giessenbachstr. 1, D-85748 Garching, Germany}

\begin{document}

\date{Accepted ???. Received ???; in original form ???}

\maketitle

\label{firstpage}

\begin{abstract}
In magnetohydrodynamics (MHD), the magnetic field is evolved by the induction equation and coupled to the gas dynamics by the Lorentz force. 
We perform numerical smoothed particle magnetohydrodynamics (\textsc{Spmhd}) simulations and study the influence of a numerical magnetic divergence.
For instabilities arising from $\divB$ related errors, we find the hyperbolic/parabolic cleaning scheme suggested by \cite{Dedner02} 
to give good results and prevent numerical artifacts from growing.
Additionally, we demonstrate that certain current \textsc{Spmhd} implementations of magnetic field regularizations give rise to unphysical 
instabilities in long-time simulations.
We also find this effect when employing Euler potentials (divergenceless by definition), 
which are not able to follow the winding-up process of magnetic field lines properly.
Furthermore, we present cosmological simulations of galaxy cluster formation at 
extremely high resolution including the evolution of magnetic fields.
We show synthetic Faraday rotation maps and derive structure functions to compare them with observations.
Comparing all the simulations with and without divergence cleaning, we are able to confirm the results of 
previous simulations performed with the standard implementation of MHD in \textsc{Spmhd} at normal resolution. 
However, at extremely high resolution, a cleaning scheme is needed to prevent the growth of numerical $\divB$ errors at small scales.
\end{abstract}

\begin{keywords}
(magnetohydrodynamics) MHD -- magnetic fields -- methods: numerical -- galaxies: clusters
\end{keywords}


\section{Introduction} \label{sec:intro}

On large scales magnetic fields are observed within galaxies, along galactic outflows \citep[see e.g.][ for a review]{2009ASTRA...5...43B}, 
within the intra-cluster medium (ICM) \citep{2006AN....327..539G} and even in the filamentary structure of the Universe \citep{2011ApJ...727L...4D}. 
Faraday Rotation Measures (RM) towards extended radio sources in and behind galaxy clusters (used to probe the magnetic field within the ICM) 
can demonstrate the turbulent nature of the magnetic field in galaxy clusters \citep{1991MNRAS.253..147T}.
Such observations, yields typically $\mu$G values for the central magnetic field within galaxy clusters, whereas 
the reconstruction of the magnetic field mainly leads to a radial decline, consistent with a turbulent magnetic field distribution 
\citep{2011A&A...529A..13K, 2010A&A...522A.105G,Vacca12, Guidetti10, Bonafede10}.

The presence of turbulence within the ICM is expected from hydrodynamical cluster simulations in a cosmological context, 
predicting a significant amount of turbulence within the ICM \citep[e.g.][]{bryan1998, 2003AstL...29..791I, rasia2004, 
dolag05, 2006MNRAS.369L..14V, 2009A&A...504...33V,2011A&A...529A..17V,2008MNRAS.388.1089I,2011ApJ...726...17P}. 
Therefore, the turbulence should amplify the magnetic field within the cluster atmosphere towards equipartition values \citep{Subramanian2006}.

Non-radiative cosmological simulations of galaxy clusters following the evolution of a primordial magnetic seed field 
have been performed using smoothed particle hydrodynamics \textsc{Spmhd} codes \citep{Dolag99,Dolag02,DolagStasyszyn09,Bonafede11} 
as well as adaptive mesh refinement (AMR) codes \citep{2005ApJ...631L..21B,DuboisTeyssier08, 2010ApJS..186..308C,2010ApJ...725.2152X, 2011ApJS..195....5M}.
Such cosmological ideal MHD simulations also predict a magnetic spectrum, efficiently redistributing the internal energy \citep{Dolag02,2005ApJ...631L..21B,Xu2009} into the ICM. 

However, to capture the features characterizing the magnetic field within galaxy clusters, 
their build-up process within the large-scale structure of the Universe has to be followed correctly.
Therefore, numerical codes have to resolve many orders of magnitude in dynamical range.
In this process, a variety of dynamo theories will be involved and tested, whereas it can point out 
towards a turbulent dynamo or shock enhancement of magnetic fields.

Here we present an extension of the standard \textsc{Spmhd} implementation in \textsc{Gadget} \citep{DolagStasyszyn09}, 
which includes divergence cleaning based on \citet{Dedner02}.
Our main focus is on simulations without radiative processes, to gain deep insight in the performance of the code itself.
Additional physical processes, would change the behaviour of the system and make it difficult to disentangle 
the possible small-scale effects resulting from a non-vanishing $\divB$ the case of our ideal MHD simulations. 
With this cleaning scheme we are able to perform extremely high resolution simulations of galaxy clusters in a cosmological context.
By reaching a spatial resolution of kiloparsec in the center of the cluster we are able to directly compare the structure of 
the magnetic field obtained in the simulations with the observed structure on the smallest scales observed.

The paper is organized as follows.
In section \ref{sec:implementation} we present the further development of the MHD \textsc{Gadget} code.
Results from typical standard tests are shown in section \ref{sec:tests}. 
In section \ref{sec:cluster} we present simulations of galaxy clusters and compare the predicted magnetic 
field structure to observations, before we summarize and conclude in section \ref{sec:conc}. 
Additionally, in the appendixes, we show more detailed the effect of dissipation and resolution on the results.


\section{SPMHD Implementation}\label{sec:implementation}

We are starting from the \textsc{Spmhd} implementation of \citep{DolagStasyszyn09} within 
the cosmological N-Body TreePM/Spmhd code \textsc{Gadget} \cite{springel01,springel05}.
For the details of the implementation we to \citet{DolagStasyszyn09} or to a more general review on \textsc{Spmhd} by \citet{Price2010}.

In short, the evolution of the magnetic field is directly followed with the induction equation.
The magnetic field acts on the gas via the Lorentz force, written in a symmetric conservative 
form using the magnetic stress tensor. Also, as originally suggested by \citet{PriceI} the fast magnetosonic 
wave velocity replaces the sound velocity within the computation of the signal velocity controlling the artificial viscosity and the time-step.
In the calculation of the gradients and divergence estimators, we follow the standard \textsc{Spmhd} implementation.

\subsection{Instability correction}

To take into account the tensile instability in \textsc{Spmhd}, which occurs when the magnetic pressure exceeds the gas pressure and 
the force between particles is becoming attractive, a correction term in the force equation is used.
This term -- introduced by \cite{2001ApJ...561...82B} and further developed in \cite{2006ApJ...652.1306B} -- 
subtracts from the equation of motion any unphysical force resulting from a non-vanishing numerical $\divB$.
Contrary to the original implementation, we restrict the correction to not exceed the Lorentz force, which is necessary at strong shocks.
Therefore, we evaluate the correction contribution and if necessary renormalise to be only as much as the Lorentz force.

\subsection{Time integration} 

In \cite{DolagStasyszyn09} the evolution of the magnetic field $\bmath{B}$ is done in physical units, 
so that in cosmological simulations the induction equation contained a term of $-2\bmath{B}$ to capture the cosmological dilution due to the expansion of space.
However, defining the magnetic field in comoving units ($\bmath{B_{c}}=\bmath{B}/a^2$), with $a$ the cosmological scale-factor 
allows to drop the $-2\bmath{B}$ term and the induction equation becomes
\begin{equation}\frac{d\bmath{B_c}}{dt}=(\bmath{B_c}\cdot\bmath{\nabla})\bmath{v}-\bmath{B_c}(\bmath{\nabla}\cdot\bmath{v}).\end{equation}
When reading output data from the simulation the magnetic field is converted back into physical units by multiplying it with $1/a^{2}$. 

To capture in more detail situations where the magnetic field structure is folded on the resolution scale, 
an additional time step criterion for every particle $i$ can be constructed:
\begin{equation}\Delta{}t_{i} \approx \frac{h_i}{v_{i}^{\mathrm{typical}}} \approx h_i\sqrt{\frac{\rho_{i}}{4\pi(\divB)_{i}^2}},\end{equation}
where $h$ is the \textsc{Spmhd} smoothing length.
This criterion allows to capture regions of high numerical divergence, where the magnetic field structures reach the resolution 
limit and put the particles on lower time-step ensuring a more detailed evolution of the magnetic field.
Although this additional criterion rarely overcomes the standard time-step criterion, there are situations 
where it seems to be quite helpful to follow the local dynamics with more details.

\subsection{Divergence Cleaning} \label{sec:ded}

It is of fundamental interest in \textsc{Spmhd} simulations to keep the magnetic divergence $\divB$ 
arising from the numerical integration schemes to a minimum.
We implement into the \textsc{Gadget} code, the divergence cleaning scheme introduced by \cite{Dedner02}, 
which evolves an additional scalar potential $\psi$ representing non-vanishing $\divB$ introduced artifacts. 
By construction, this potential $\psi$ propagates the numerical errors outwards the simulation, while damping them, 
by subtracting the gradient of $\psi$ in the induction equation. 
This method is also widely used in Eulerian codes \citep[i.e.][]{Cecere2008,Anderson2006,2012JCoPh.231..718K} and 
recently introduced in the moving mesh code \textsc{Arepo} \citep{Springel2010,Pakmor2011}.
First attempts to use this technique in \textsc{Spmhd} were made by \cite{PriceIII}, being not satisfactory.
They found only a mild improvement of the numerical $\divB$ errors and in some test cases 
the cleaning scheme was even causing instabilities particularly in 3D.
The implementation shown here, lowers the $\divB$ error in general, does not show unwanted effects and 
additionally leads to a very small numerical diffusion.

Following \cite{Dedner02}, therefore assuming a non-vanishing $\divB$, an additional term entering the induction equation can be derived:
\begin{equation}\left.\frac{d\bmath{B}}{dt}\right|_{i}^{\rmn{Ded}}=-(\nabla\psi)_{i}.\end{equation}
To be energy conserving, the removed of magnetic energy is transferred into internal energy or entropy $A$ at a rate of
\begin{equation}\left.\mu_0\frac{dA}{dt}\right|_{i}^{\rmn{Ded}}=-\frac{\gamma-1}{\rho_{i}^{\gamma-1}}\bmath{B}_{i}\cdot(\nabla\psi)_{i}\end{equation}
where $(\gamma-1)/\rho^{\gamma-1}$ is the conversion factor from internal energy to entropy with the adiabatic index $\gamma$.

However, the scalar potential $\psi$ has to be chosen in a way to actually removes numerical errors.
\cite{Dedner02} found that the most effective solution is to construct and evolve $\psi$ propagating 
the errors away from the source (i.e hyperbolic cleaning) and damping them (i.e parabolic cleaning).
This results in the following evolution equation for $\psi$:
\begin{equation}\frac{d\psi_{i}}{d{}t} = -\left((c_\rmn{h})_{i}^2(\divB)_{i} - \frac{\psi_{i}}{\tau_{i}}\right)\end{equation}
which shows that $\psi$ now satisfies a wave equation propagating the errors outwards the source with a speed of $c_\rmn{h}$ (first term of the equation) 
and decaying them on a timescale of $\tau$ (second term in the equation).
It is again natural in \textsc{Spmhd} simulations to relate the propagation speed to the fast magnetosonic wave, hence using $c_\rmn{h}=\sigma{}v^\rmn{mhd}$.
Note, that this velocity does not have to be related with any special quantity per se, even can be a constant value.
Also the timescale $\tau$ can be related to a typical length scale (smoothing length $h$) and 
velocity resulting in $h/\lambda{}v^\rmn{mhd}$, only leaving dimensionless numerical constants $\lambda$ (parabolic),
$\sigma$ (hyperbolic) of order unity.
We choose values of $\lambda=4$ and $\sigma=1$ to recover the best solution in \cite{PriceIII}.

Similar to the tensile instability correction of the Lorentz force \citep{2001ApJ...561...82B} in the equation of motion,
this method can lead to instabilities. In particular, situations where small scale structures in the magnetic field 
lead to an imprecise calculation of the $\divB$ source term. 
This situation will manifest as an over-correction of the induction equation. 
Hence, we use a limiter for the cleaning contribution in a similar form, not allowing the correction in the induction equation 
to be larger than a given value $Q$ weighted by the local induction value.
When the correction exceeds the original term, we renormalise it.
To ensure stability, this ratio $Q$ has to be less or at most equal to $1$.
Testing different parameters we found a value of $Q=0.5$ to be sufficient in ensuring a proper evolution of the magnetic field
while avoiding over-corrections due to the cleaning scheme.

Note, that \cite{Pakmor2011} use a global maximum for the signal velocity in the evolution of the scalar field $\psi$.
However, our \textsc{Spmhd} formulation of the cleaning equations takes all quantities per particle, in order to avoid any 
dissipation or overcorrection in already stable regions.


\section{Application to test problems}\label{sec:tests}

The problems in structure formation are very complex and astrophysical objects of interest evolve in a strongly, 
non-linear way from the initial conditions to the final stages during the different cosmological epochs.
To be confident about the numerical results, the hydrodynamical solving scheme has to be tested properly and compared with known analytical solutions.
Therefore, we tested the new cleaning scheme in an extensive series of shock tubes and planar tests, similar as done in \cite{DolagStasyszyn09}. 
In the same way as done previously, we performed all the tests by setting up a fully three-dimensional glass like particle distribution, 
to obtain results under most realistic possible circumstances and compare then with the solution obtained with \textsc{Athena} \citep{Stone2008} (in 1D or 2D, respectively).
Additionally, the $\divb$ errors are defined by the dimensionless quantity
\begin{equation}   \mathrm{Err}_{\divb_i} = \vert \divB_i \vert \frac{h_i}{|\Bf_i|} \label{eq:divbe}\end{equation}
which can be calculated for each particle, and determines the reliability of the results. 

\subsection{Shock-tube tests}

The most common MHD shock tube test is the \citet{BrioWu88} shock \citep[Test 5A in][]{1995ApJ...442..228R}.
In this shock-tube, a shock and a related rarefaction are moving together.
For the standard scheme, we used the method described in \cite{DolagStasyszyn09}. 
Additionally, the instability correction as described in section \ref{sec:implementation} was used.
The new \textsc{Spmhd} results generally agree with the solution obtained with \textsc{Athena}, 
although there is some residual scatter in the individual particle values within the 3D volume elements, 
as well as some small scale noise, especially in the low-density part.
Note, that the mean values for the internal energy, the velocity or the magnetic field, 
can locally show some minor but systematic deviations from the ideal solution.
However, the total energy shows much better, nearly unbiased, behaviour. 
This demonstrates the conservative nature of the symmetric formulations in \textsc{Spmhd}.

\begin{figure}
  \begin{center}
  \includegraphics[width=0.45\textwidth]{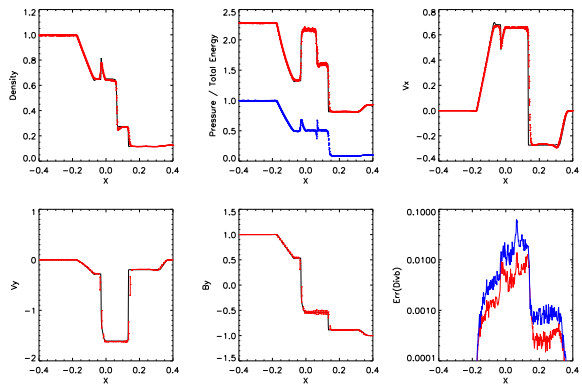}
  \end{center}
  \caption[]{Shocktube 5A at time $t=7$ with the Dedner cleaning scheme. 
  In the first row is the density (left panel), total energy and pressure (middle panel) and the $x$ velocity component (right panel). 
  The second row shows the $y$ velocity component (left panel), the $y$-component of the magnetic field (middle panel) and the measure of the $\divb$ error (right panel). 
  There we also show the $\divb$ error of the Standard MHD implementation (in blue). 
  The colored lines with error bars (blue for the pressure in the upper middle panel) show the \textsc{Spmhd} results, 
  the black lines are reference results from \textsc{Athena} in a 1D setup.}
  \label{fig:BrioWu_Ded}
\end{figure}

As can be seen in \FIG{BrioWu_Ded}, the cleaning scheme achieves a visible reduction of the $\divb$ error and lowers the numerical noise.
To find good numerical parameters for the cleaning scheme, we performed a parameter study test series.
This study takes into account the estimated noise and accuracy, as previously explained in \cite{DolagStasyszyn09}.
The best results are obtained for parameters ranging from $\lambda=2\sim 5$ to $\sigma \sim 1$.
We therefore have chosen the values of $\lambda=4$ and $\sigma=1$, similar as used by \cite{PriceIII}.
These numbers imply (see Section \ref{sec:ded}) that the $\divb$ errors propagate at the signal velocity 
and will be damped within approximately $2 \sim 5$ smoothing lengths distances.

\begin{figure}
  \begin{center}
  \includegraphics[height=0.4\textwidth]{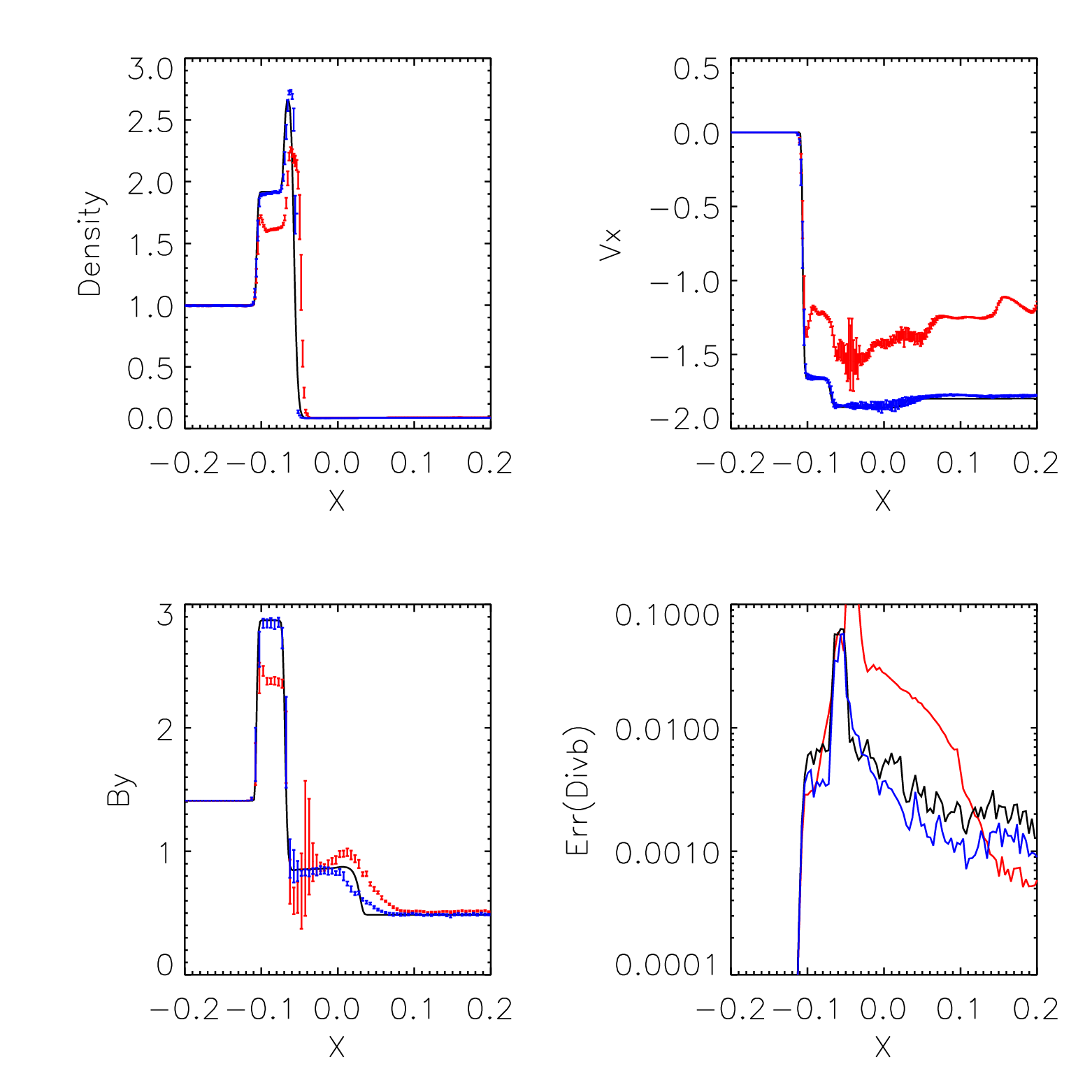}
  \end{center}
  \caption[]{Shock-tube 1B from \citet{1995ApJ...442..228R}. 
  In black are the result from \textsc{Athena}, in blue our cleaning scheme with $Q$ delimiter and in red without. 
  From upper left to bottom right, density, $x$ velocity, $y$ magnetic field and $\divb$ errors are plotted. 
  Again, in the last panel, are $\divb$ errors for the Standard case. 
  With the limiter $Q\le0.5$ the cleaning scheme leads to less $\divB$ error than the standard MHD implementation (black line in the lower right panel).}
  \label{fig:Shock1B}
\end{figure}

\FIG{Shock1B} shows a case, where the original cleaning scheme without limiter fails. 
However, with the limiter $Q\le0.5$, no instability can be seen and we notice 
the $\divB$ error again to be smaller than with the Standard implementation. 
Comparing results of different regularization schemes as presented in \cite{DolagStasyszyn09} 
and the cleaning scheme presented here, we note that the cleaning scheme does not smooth sharp 
structures within the different shock-tube tests, as other regularization methods described in \cite{DolagStasyszyn09} are doing.

To check the performance of the new implementation in detail, we ran the full set of different shock-tube tests as presented in \citet{1995ApJ...442..228R}.
The results are summarized in \FIG{comp_imp}, where the black diamonds represent the results of the 11 shock-tube test.
The cleaning scheme is stable for all 11 shock-tube tests and the $\divB$ values decrease on average by a factor $2\sim3$.

\subsection{Planar Tests}

Besides the Shock-tube tests described in the previous section, two dimensional (e.g. planar) test problems are an excellent test-bed 
to check the performance of MHD implementations.
Such higher dimensional tests include additional interactions between different evolving components with non-trivial solutions. 
They can be quite complex (with several classes of waves propagating in several directions) such as the Orszag-Tang Vortex 
or simple (but with strong MHD discontinuities) like the Strong Blast or Fast Rotor.

\subsubsection{Fast Rotor}

\begin{figure*}
  \begin{center}
  \subfigure[Fast Rotor - \textsc{Athena}]{
  \label{fig:rot_ath}
  \includegraphics[height=0.4\textwidth]{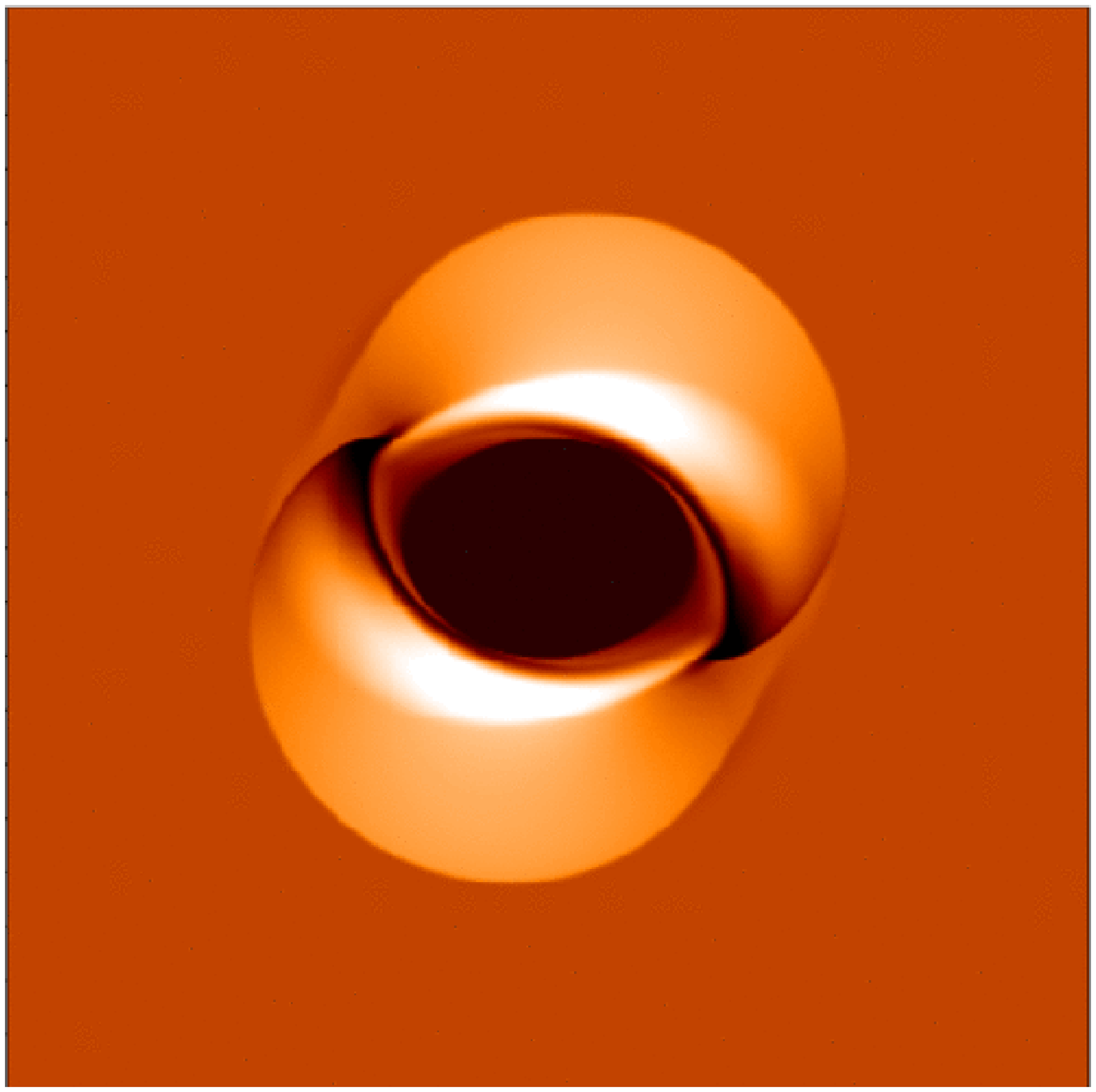}}
  \subfigure[Fast Rotor - Dedner]{
  \label{fig:rot_ded}
  \includegraphics[height=0.4\textwidth]{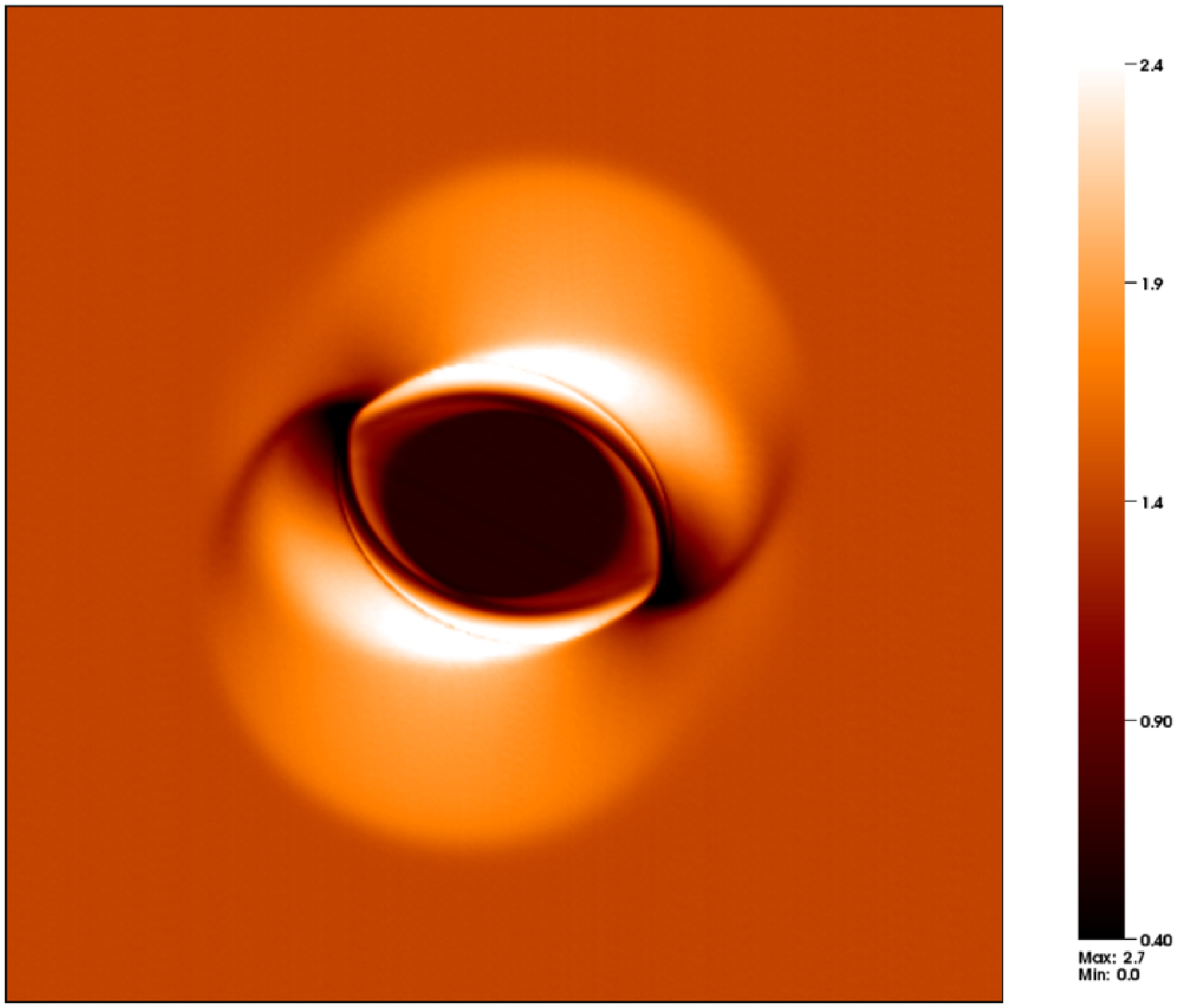}}
  \end{center} 
  \caption[]{Magnetic field strength in the Fast Rotor test at $t=0.1$. 
  The \textsc{Athena} solution of the test problem is shown in the left panel. 
  In the right panel is the result obtained with the \textsc{Spmhd} Dedner scheme. 
  All the main features are reproduced. The shape, positions and amplitudes correspond well, 
  although the \textsc{Spmhd} run shows slightly more smoothing, especially at the outer boundary.}
  \label{fig:Rotor}
\end{figure*}

The Fast Rotor test was first introduced by \citet{Balsara99}, to study star formation scenarios, in particular the strong torsional 
Alfv\'en waves and is also commonly used to validate MHD implementations \citep[for example see][]{Toth00,LondrilloDelZanna00,PriceIII,2006ApJ...652.1306B, DolagStasyszyn09}.
The test consists of a fast rotating dense disk embedded in a low density, static and uniform medium, with an initial constant magnetic field along the x-direction 
(e.g. $B_x=2.5\pi^{-1/2}$). 
In the initial set-up, a disk with radius $r=0.1$, density $\rho=10$ and pressure $P=1$ is spinning with an angular velocity $\omega=20$, 
embedded in an uniform background with $\rho = P =1$.
Again we used a glass like 3D particle distribution with $700\times700\times5$ particles and periodic boundaries, 
and increased the amount of particles in the disk to achieve the desired density using the same mass per particle. 
For comparison, we obtained a reference simulation with \textsc{Athena} run at $400\times400$ cells.
The results obtained with the Dedner scheme and with \textsc{Athena} are shown in \FIG{Rotor}.
The shape, positions and amplitudes correspond very well, although the \textsc{Gadget} runs appears slightly more smoothed.
Here, the cleaning scheme reduces the overall numerical $\divB$ error by a factor of two, as can be seen from the according data point in Fig. (\ref{fig:comp_imp}).

\subsubsection{Strong Blast}

\begin{figure*}
  \begin{center}
  \subfigure[Strong Blast - \textsc{Athena}]{
  \label{fig:Blast_ath}
  \includegraphics[height=0.4\textwidth]{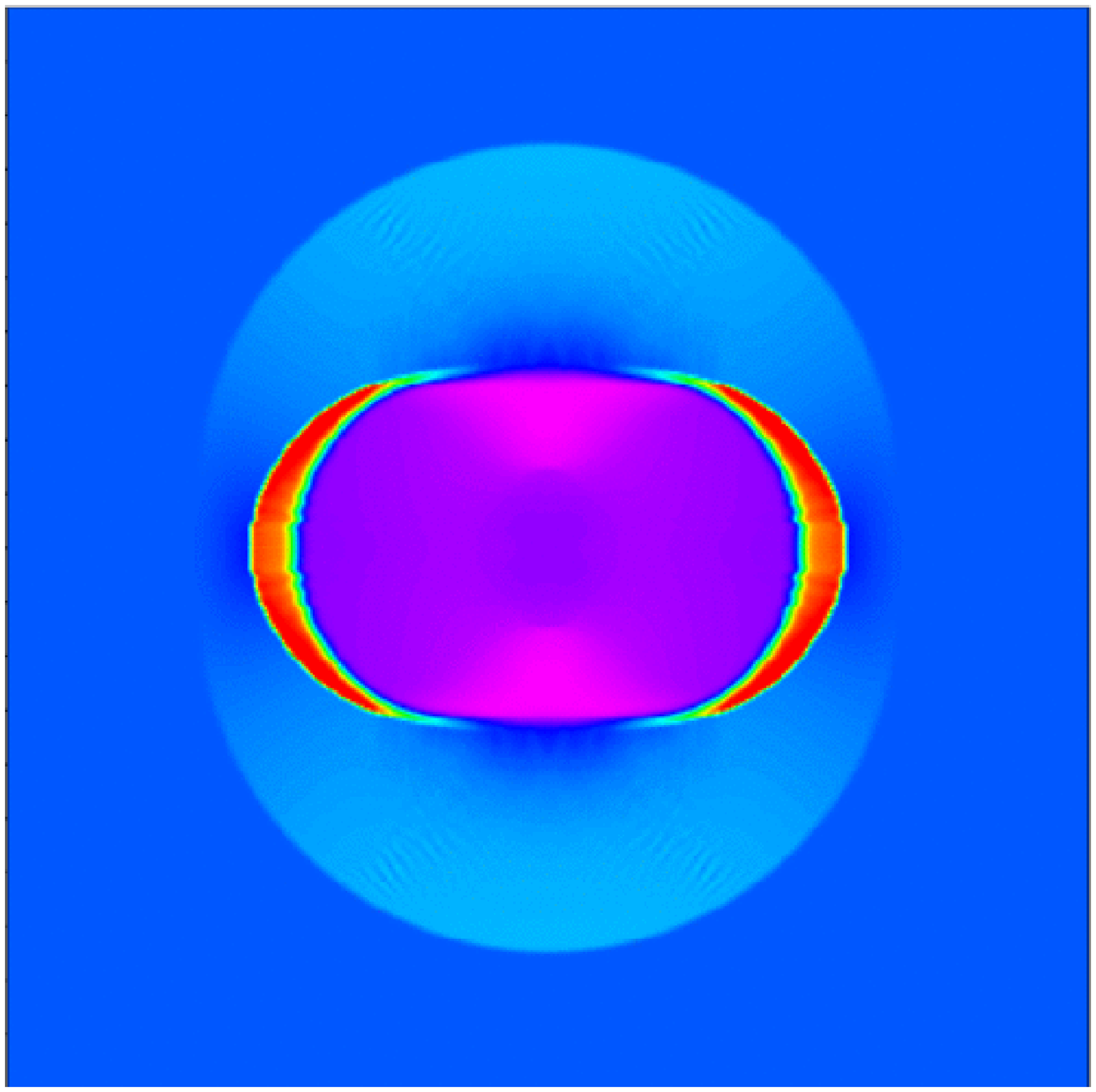} }
  \subfigure[Strong Blast - Dedner]{
  \label{fig:Blast_ded}
  \includegraphics[height=0.4\textwidth]{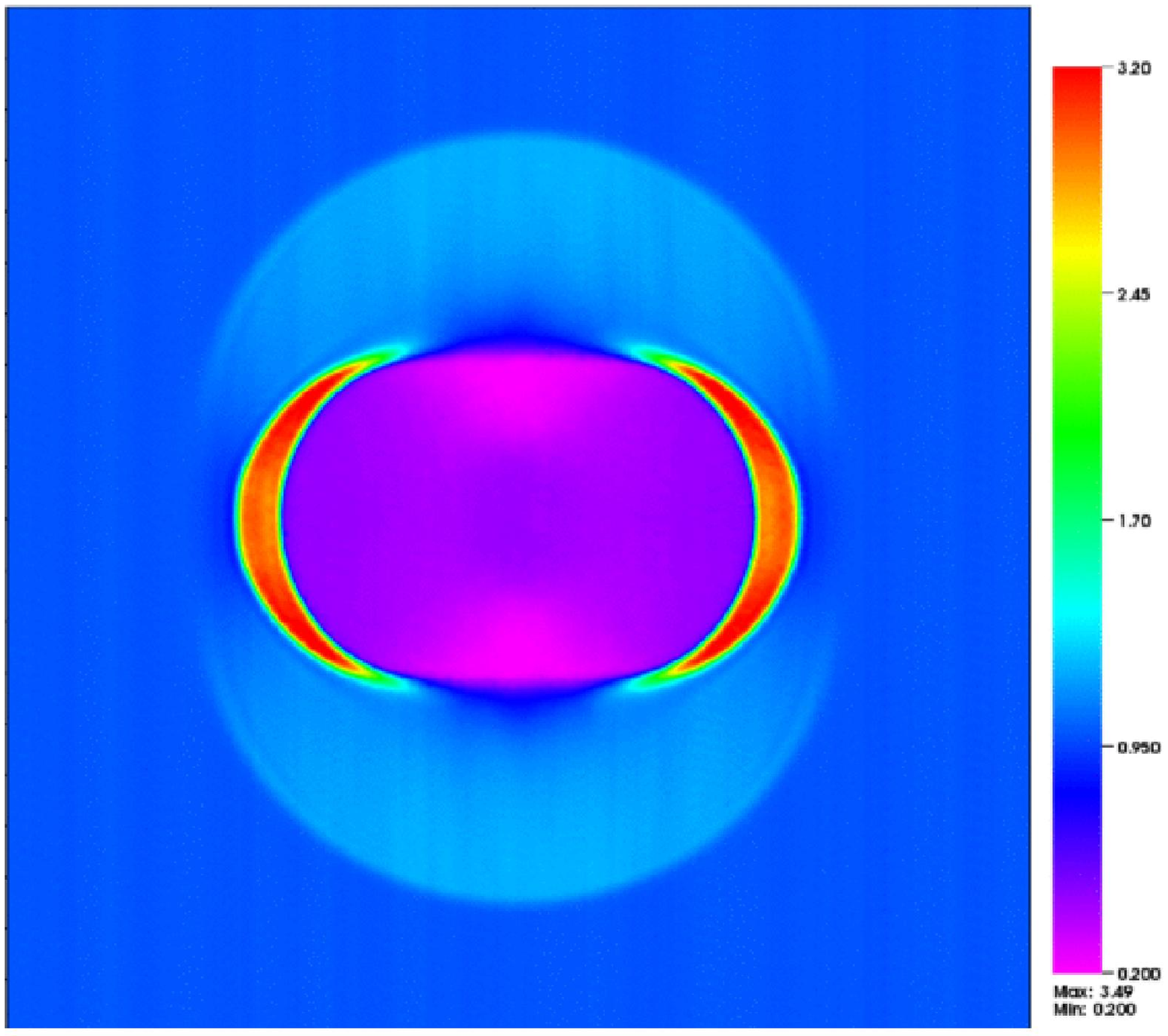} }
  \end{center} 
  \caption[]{Density distribution for the Strong Blast test at $t=0.02$. 
  The left panel shows the \textsc{Athena} results, while the right panel shows the \textsc{Spmhd} Dedner scheme. 
  There is no visible difference between both plots, beside the small intrinsic smoothing of sharp features within the \textsc{Spmhd} solution.}
  \label{fig:Blast}
\end{figure*}

The Strong Blast test is of an explosion of a circular hot gas blob within a static magnetized medium and is also 
commonly used for MHD code validation \citep[see for example][]{LondrilloDelZanna00,Balsara99}.
It consist of a background medium with a constant density $\rho=1$, where a hot disk of radius $r_0=0.125$ and 
pressure $P_d=100$ is embedded in gas with a pressure of $P_0=1$.
Additionally, there is initially an homogeneous magnetic field in the $x$-direction, with a strength of $B_x=10$.
The system is evolved until $t=0.02$ and an outgoing shock wave develops, travelling not circular, but along the magnetic field lines.
\FIG{Blast} shows the density at the final time, comparing the \textsc{Athena} results with the Dedner scheme in \textsc{Gadget}. 
There is no visible difference between the \textsc{Spmhd} implementations and the \textsc{Athena} results. 
Besides some very small variations, there is no significant difference between the various \textsc{Spmhd} schemes, and all features are well reproduced.
In this test problem, the $\divB$ cleaning scheme has effectively a similar dissipative behavior than the scheme with artificial dissipation.
As can be seen from the according data point in Fig. (\ref{fig:comp_imp}), 
the cleaning as well as the other regularization methods give moderate improvements in the numerical $\divB$ errors.  

\subsubsection{Orszag-tang Vortex}

\begin{figure*}
  \begin{center}
  \subfigure[Orszag-Tang Vortex- \textsc{Athena}]{
  \label{fig:vor_ath}
  \includegraphics[height=0.4\textwidth]{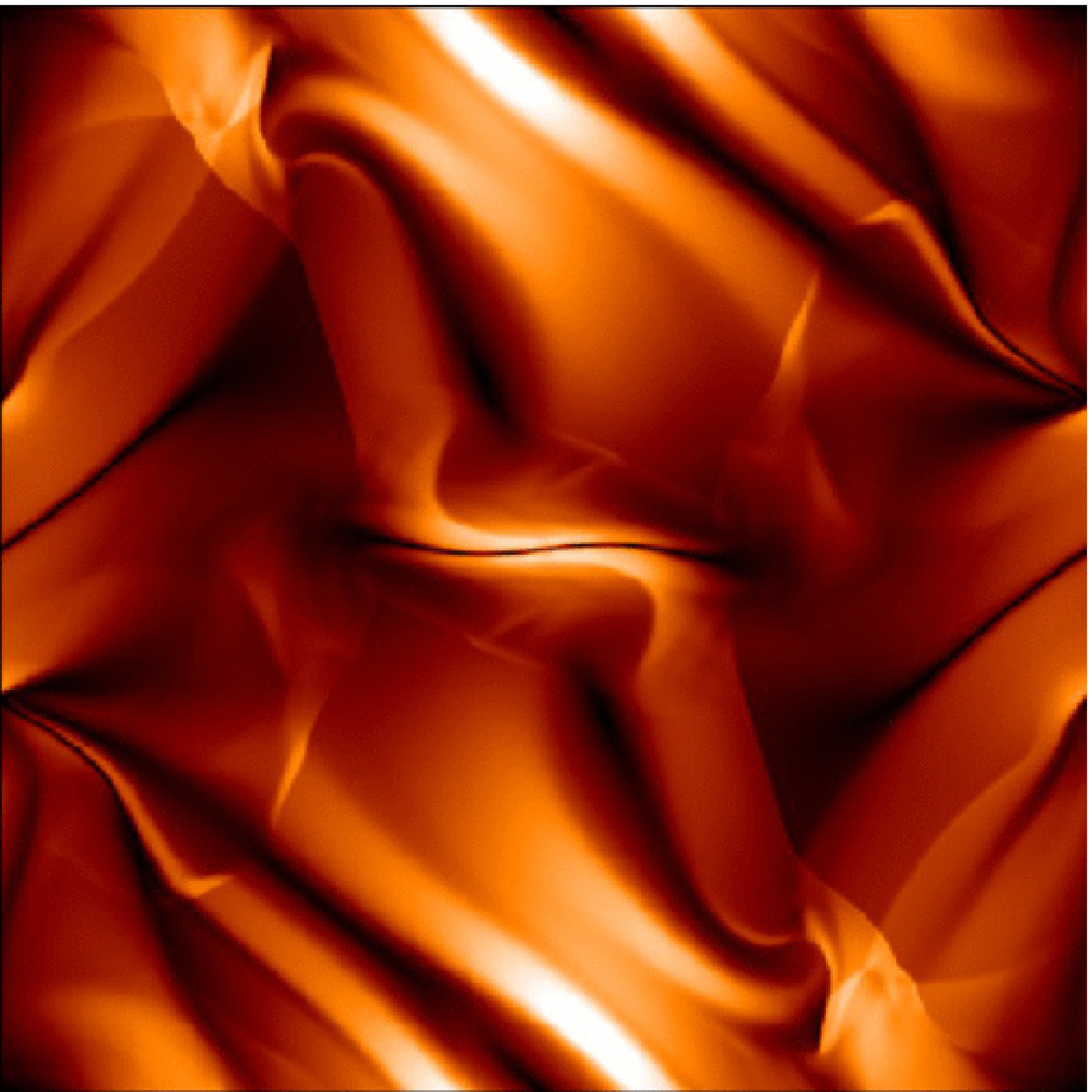} }
  \subfigure[Orszag-Tang Vortex- Dedner]{
  \label{fig:vor_ded}
  \includegraphics[height=0.4\textwidth]{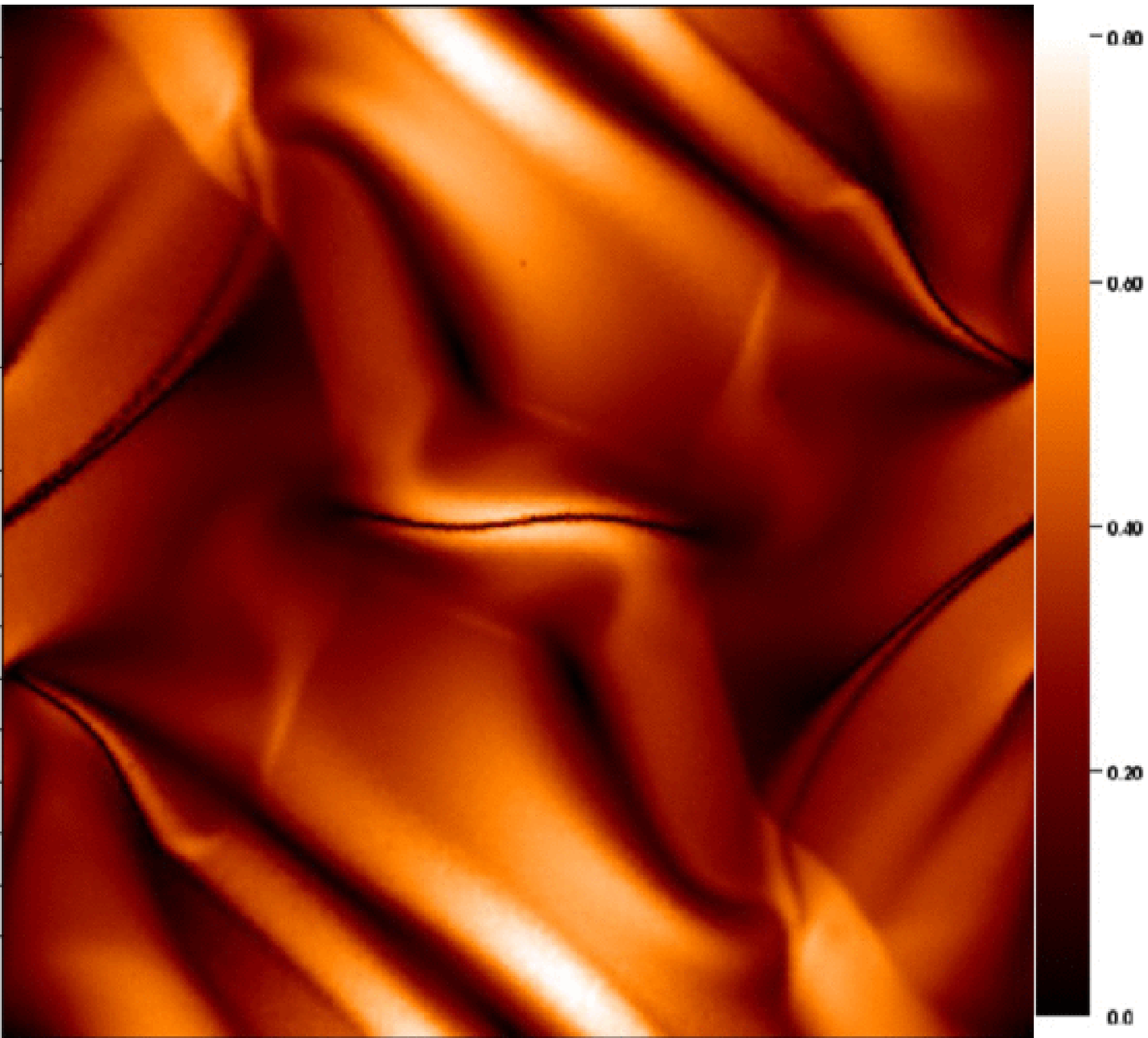}}
  \end{center} 
  \caption[]{The magnetic field strength in the Orszag-Tang Vortex at $t=0.5$. The left panel shows the \textsc{Athena} results, 
  while the right panel shows the \textsc{Spmhd} Dedner scheme. Again, some the sharp features are slightly smoothed in the \textsc{Spmhd} 
  implementations but overall the results compares very well.}
  \label{fig:Vortex}
\end{figure*}

\begin{figure}
  \begin{center}
  \includegraphics[width=0.4\textwidth]{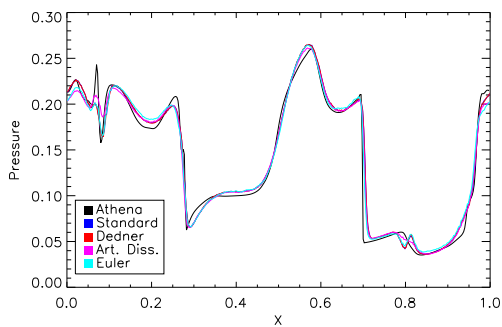}
  \end{center} 
  \caption[]{A $t=0.5$ cut of the pressure in the Orszag-Tang Vortex at $y=0.3125$. 
  The black line reflects the results obtained with \textsc{Athena}, while the colors show different MHD schemes. 
  The cut can be compared with existing results, e.g. \citet{2006ApJ...652.1306B}. 
  Note, that the variations in the solution between schemes mainly result from different dissipative characteristics. 
  However, the overall solutions are in good agreement.}
  \label{fig:VortexCut}
\end{figure}

This planar test problem, introduced by \citet{Orzang79}, is well known for the interaction between several classes of shock waves 
(at different velocities) and the transition to MHD turbulence.
Also, this test is commonly used to validate MHD implementations \citep[for example see][]{1994JCoPh.115..485D,1991PhFlB...3...29P,LondrilloDelZanna00,PriceIII,2006ApJ...652.1306B}.
It consists of ideal gas with $\gamma=5/3$ within a a box of (e.g. $x=[0,1],y=[0,1]$) and periodic boundaries conditions.
The velocity field is defined by $v_x=-\sin(2\pi y)$ and $v_y=\sin(2\pi x)$.
The initial magnetic field is set to $B_x=B_0 v_x$ and $B_y=B_0 sin(4\pi x)$. 
The initial density is $\rho=\gamma P$ and the pressure is $P=\gamma B_0^2$. 
An usual time to evaluate the system $t=0.5$.
\FIG{Vortex} shows the final result at that time for the magnetic pressure for the \textsc{Athena} run (left panel), 
and the Dedner scheme (right panel).
The results are quite comparable, however, the use of \textsc{Spmhd} method leaves its imprint in a slightly smoothed appearance in the \textsc{Gadget} results.
This can also be seen in \FIG{VortexCut}, which shows a cut through the test for different implementations, 
comparable with other cuts done in the literature \citep{2006ApJ...652.1306B}. 
In general there is reasonable agreement, however all the \textsc{Spmhd} results clearly show a smoothing of some features. 
However, the Dedner and standard implementations, tend to match better some regions that the dissipative schemes oversmooth 
(e.g. region near $x\sim1$ in \FIGp{VortexCut}), even better that the Euler scheme.
Note, that a exact comparison is difficult, mainly because this test includes the propagation of several types of magnetosonic waves, 
implying that if missing the correct velocity (i.e. by some dissipative effect) of a particular wave, the result will diverge between implementations.

This periodic test in particular, is good for checking the \textsc{Spmhd} implementation based on the Euler potentials formalism, 
finding a very good agreement with other authors \citep[i.e.][]{Rosswog07}. 
As the Euler potentials are be $\divB$ free by construction, any numerical arising $\divB$ error can clearly traced back to the numerical 
inaccuracies in \textsc{Spmhd} formalism itself.
Our major interest in this scheme is therefore the possibility to measure the errors that arise from the interpolation.

\begin{figure}
  \begin{center}
  \includegraphics[width=0.45\textwidth]{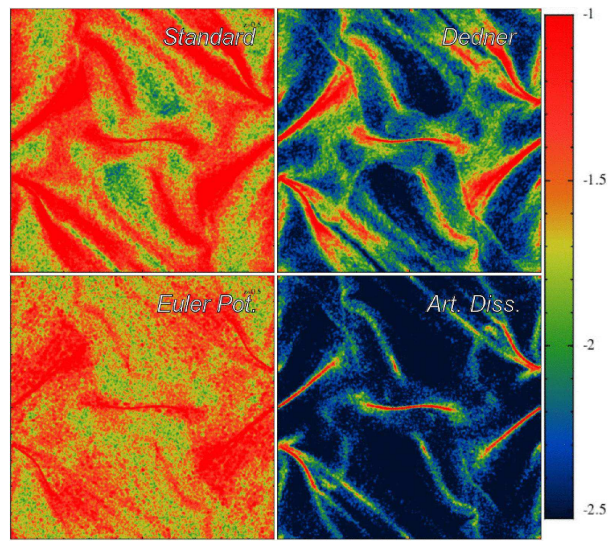} 
  \end{center}
  \caption[]{$\divb$ errors as defined by \EQ{divbe} in the Orszag-Tang Vortex at $t=0.5$. 
  The upper row shows the standard scheme (left panel) and the Dedner scheme (upper right) panel. 
  The lower row shows the calculated $\divb$ errors from the Euler potential formulation (left panel) and a run with artificial dissipation (right panel). 
  The Euler potential runs defines the real numerical limits of the simulation and confirms the need of a numerical cleaning scheme.}
  \label{fig:VortexDivbe}
\end{figure}

In \FIG{VortexDivbe} we show the calculated $\divb$ errors as defined by \EQ{divbe}. 
It can be seen that the numerical errors in the standard \textsc{Spmhd} implementation are only slightly 
larger than the errors of the implementation based on Euler potential, however the spacial distribution varies.
This is the numerical error limitation, which can be overcome by using higher resolution.
Additionally, the numerical $\divB$ errors we see are caused by the magnetic field structures getting folded below the kernel scales.
Then, the basic assumption on which \textsc{Spmhd} works, namely that the values of any quantity of interest are smooth below the kernel scales, starts to get violated.
In Eulerian methods such structures are automatically mixed (e.g. dissipated) on the resolution scale.
In \textsc{Spmhd} an extra scheme (i.e artificial dissipation) is needed to remove those small scale structures, 
acting as regularization of the field below the kernel scale.
The Dedner cleaning scheme acts in this way, however, as seen already in previous test, it dissipates mainly the magnetic field 
structure below the kernel scale and does not lead to strong smoothing of the field on scales larger than the kernel scale. 
However, is enough to significantly remove the numerical $\divB$ errors, as shown in \FIG{VortexDivbe}.
More drastic approaches suffer from the same issue, as can be seen for the artificial dissipation case.

\subsubsection{Long time stability of the Orszag-tang Vortex}

\begin{figure*}
  \begin{center}
  \includegraphics[width=0.95\textwidth]{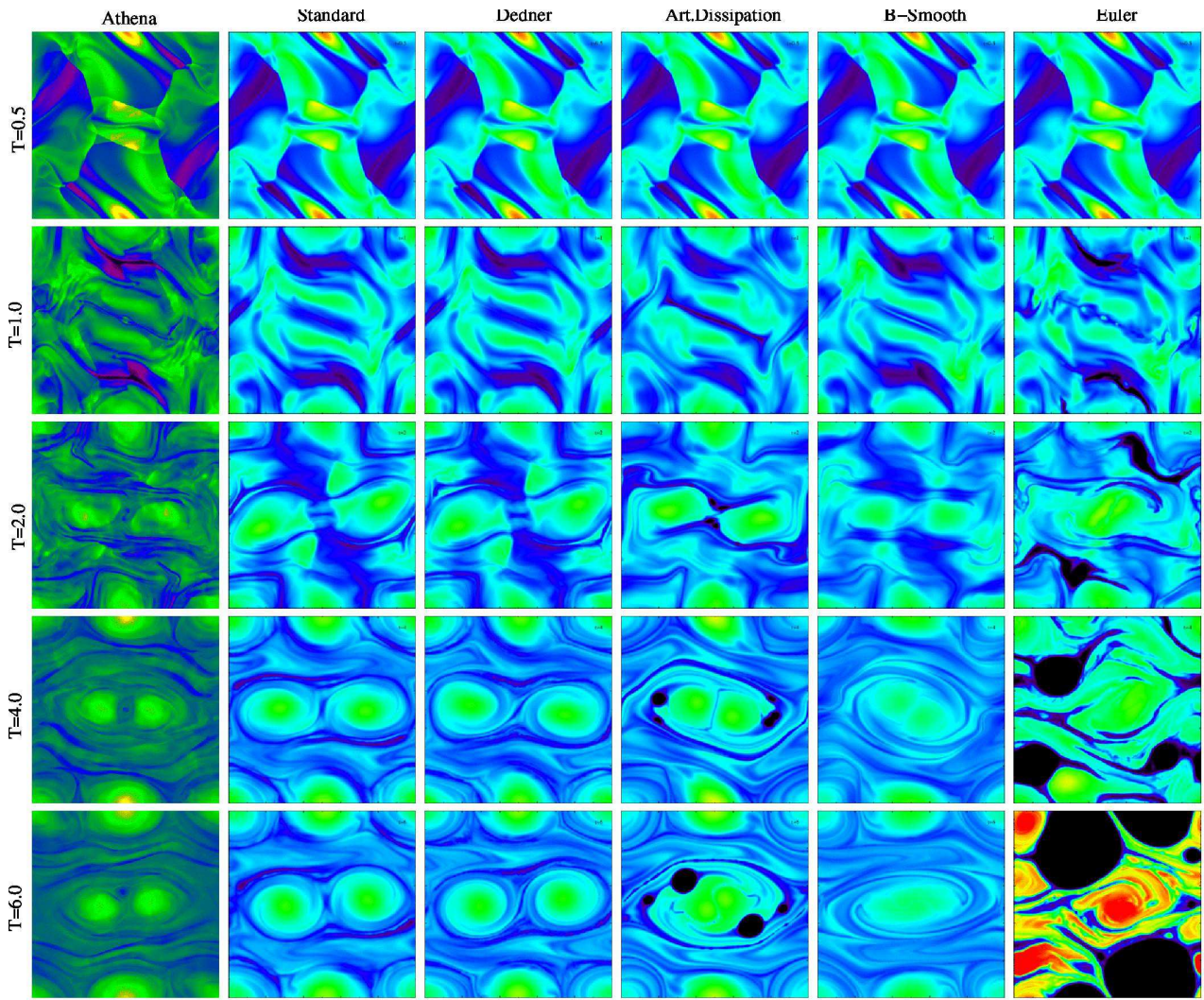}
  \label{fig:vor_long}
  \end{center} 
  \caption[]{Density maps of the Orszag-Tang vortex at various times $T=0.5,1,2,4,6$ using several MHD schemes. There are deviations from the expected solution (\textsc{Athena}), even appearing at early times $T=1.0$ for Euler, and later in the more dissipative schemes. There are also problems long-time simulations.}  
  \label{fig:Vortexlong}
\end{figure*}

The long time evolution of the Orszag-Tang Vortex was first studied by \citet{1991PhFlB...3...29P}.
They focused on the possible stable solutions of the supersonic flow.
Therefore, they studied the long time evolution varying initial Mach numbers.
Interestingly, they always found a quasi stable configurations in the long time evolution, where they typically evolved the problem until $t=8$.
We run the Orszag-Tang vortex to large times to investigate the stability of the different implementations 
as well as the influence of the differently strong numerical dissipation.
Additionally, we wanted to confirm the limitations of the Euler potential formalism \citep{2010MNRAS.401..347B}. 
In Fig. \ref{fig:Vortexlong}, we show the long time evolution of the density distribution at various times.
As expected, the implementation based on Euler potentials start to deviate from the expected solution quite early ($t=1$) 
and even runs into some severe instability at larger times $t>4$. 
The regularization scheme based on artificial dissipation, as well as periodically smoothing the magnetic field, 
show some significant effects of the underlying dissipation at times $t>4$. 
These deviations even develop an instability within the scheme based on artificial dissipation, and even earlier times $t<4$ for higher dissipation constants. 
However, both the standard MHD implementation as well as the one based on the Dedner cleaning scheme show excellent performance in the long term evolution and stability, 
well comparing to the results presented in \citet{1991PhFlB...3...29P}.
It is worth to note that both schemes appear to have less numerical dissipation than \textsc{Athena} 
(See http://www.astro.princeton.edu/~jstone/athena.html for a comparison run until $T=5$).
In our \textsc{Athena} run, the two central density peaks start to approach each other, while
they are still stable in both, the standard as well as the Dedner \textsc{Spmhd} implementation.

\subsection{Discussion}\label{sec:discuss}

\begin{figure*}
  \begin{center}
  \includegraphics[width=0.75\textwidth]{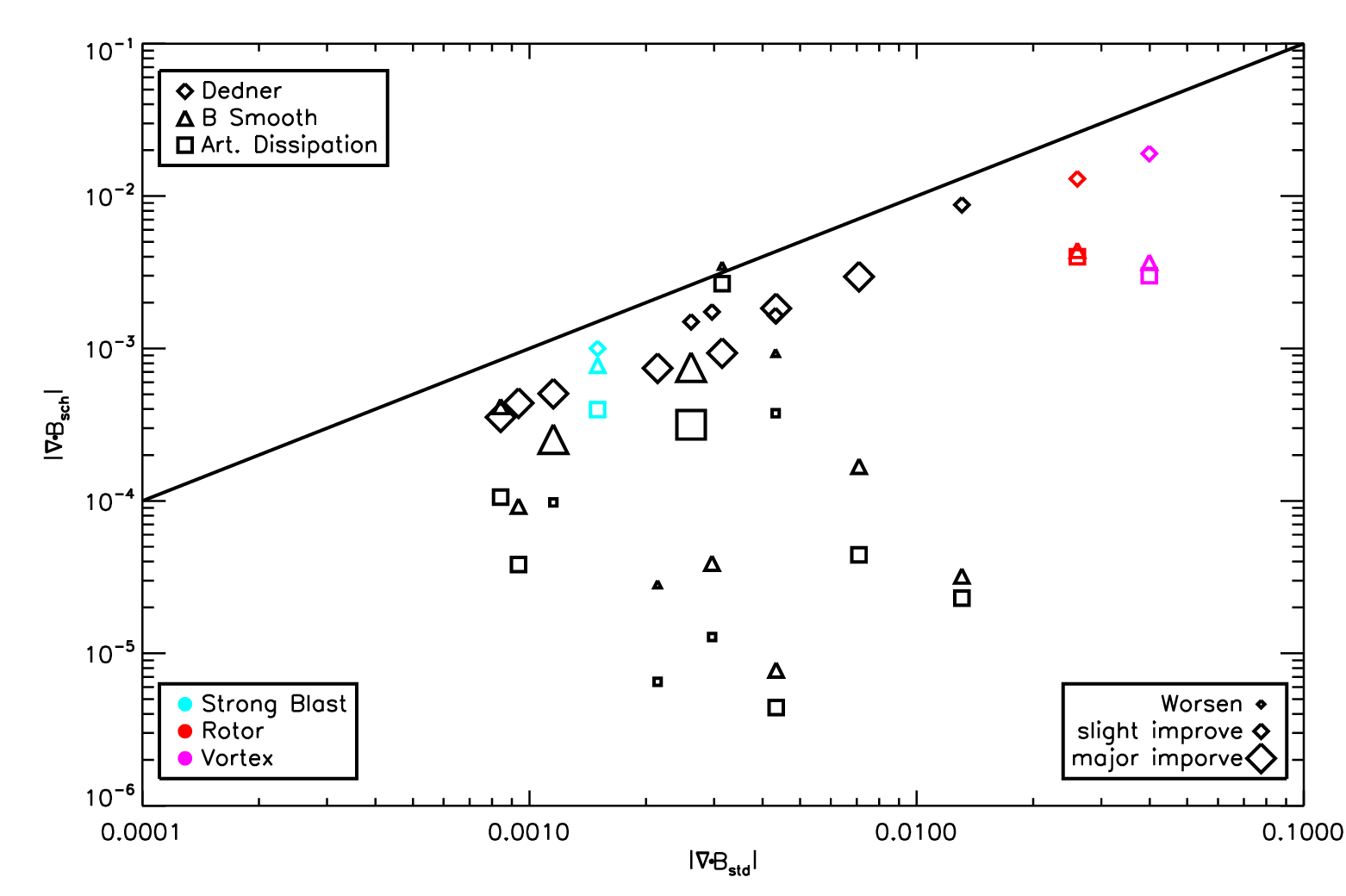}
  \end{center} 
  \caption[]{Summary plot showing the results of all 1D and 2D tests for the different \textsc{Spmhd} implementations, 
  comparing the average $|\divB|$ errors from the different schemes (symbols) and the Dedner $\divB$ cleaning against the standard \textsc{Spmhd} implementation. 
  In black are the shock tube tests and in colors the advanced problems. 
  The points sizes follows a quality criterion defined in \citet{DolagStasyszyn09}, comparing the \textsc{Spmhd} solutions with the Euler ones. 
  We define 3 different sizes when compared to the standard quality value, corresponding to worst (small), major (big) or minor (normal) improvement.}
  \label{fig:comp_imp}
\end{figure*}

In \FIG{comp_imp} we show the comparison of the performance of the different \textsc{Spmhd} implementations among all the tests discussed in the previous sections.
We compare the numerical errors obtained by the different implementations against the standard implementation.
Additionally, we re-size the points, following the quality criterion as defined in \citet{DolagStasyszyn09}. 
Therefore we are able to compare the performance between all against the standard runs.
For this, we define 3 different sizes when comparing to the standard quality value, corresponding to different grades in the improvement.

The Dedner cleaning schemes clearly reduced the $\divb$ error in all tests presented. 
Although the dissipative schemes have even lower $\divb$ errors, the cleaning scheme out stands, as it lowers the $\divb$ and also is least dissipative.
In contrast, there some tests showing lower $\divb$ errors, but the regularization schemes over smooth some features, 
therefore enhancing differences with the correct solution. 

Especially regularization schemes like the one based on artificial dissipation can be seen to mimic the Ohmic dissipation and therefore leading away from ideal MHD.
In galaxy clusters, such effects might have to be taken into account to obtain results which better agree with the observed profile of the resulting magnetic fields 
\citep{Bonafede11}, and therefore it is quite important to have an underlying MHD implementation which does not suffer from any form of artificial dissipation.
To study the structure of the magnetic field as imprinted by the complex, hydro-dynamical flows as imprinted by structure formation, 
it is quite important to have a scheme which on one hand regularizes the magnetic
field below the kernel scale to avoid unwanted numerical artifacts as well as not influencing the magnetic field structure at scales larger than the kernel scale.
As shown by the tests performed in the previous section, the \textsc{Spmhd} implementation based on the Dedner $\divb$ cleaning scheme seems to fulfill these requirements.


\section{Galaxy cluster and magnetic fields}\label{sec:cluster} 

In the hierarchical picture of structure formation, small objects collapse first and then merge to subsequently form larger structures.
This formation process is reflected in the intricate structure of galaxy clusters, which properties depend on how the structure of the 
smaller objects was accreted by the cluster and its gravitational potential.
These accretion or merging events cause shocks and turbulence in the ICM, leading to a redistribution or amplification of magnetic fields.
To describe this process within numerical simulations is extremely challenging, as the structures in and around galaxy clusters are intricate 
and range over many dynamical scales and orders of magnitude.
Here, we used the so called Zoomed Initial Conditions (ZIC) technique \citep{1993ApJ...412..455K,tormen97} to follow the evolution of a single, 
relatively low mass galaxy cluster within a large cosmological box with so far unreached precision to study the evolution and the final structure 
of the magnetic field and compare it with the observed magnetic field structure within a similar, low mass galaxy cluster.

To perform this study, we re-simulated a Lagrangian region selected from a lower resolution dark matter only cosmological box.
This parent simulation has a box--size of $684$ Mpc, and assumed a flat $\Lambda$CDM cosmology with $\Omega_m=0.3$ (matter density), 
$h=0.7$ (Hubble constant), $f_{bar}=0.13$ (baryon fraction) and $\sigma_8=0.9$ (normalization of the power spectrum). 
The selected cluster with a final mass of $1.5\times10^{14}~M_\odot$ was re-simulated using 4 different particle masses for the high resolution region.
To optimize the setup of the initial conditions, the high resolution region was sampled with a grid of $16^3$ cells, 
where only sub-cells were re-sampled at high resolution to allow for quasi arbitrary shapes. 
The exact shape of each high-resolution region was iterated by repeatedly running dark-matter only simulations, 
until the targeted objects were clean of any lower-resolution boundary particle out to $3\sim 5$ virial radii. 
The initial particle distributions, before adding any Zeldovich displacement, were taken from a relaxed glass configuration \citep{1996clss.conf..349W}.
The highest-resolution (e.g. 130x) thereby corresponds to a mass of the dark matter and gas particles of $1.2\times10^7 M_\odot$ and $2.3\times10^6 M_\odot$ respectively.
The according gravitational softening is $1.6~\kpc$.
Table \ref{tab:0} summarizes the settings for the different resolutions.
For simplicity we assumed an initially homogeneous magnetic field seed of $10^{-11}$ G.
Additionally, we ran several of the low-resolution versions of the simulations with our different \textsc{Spmhd} 
schemes to check the influence of the different techniques on the results.
To focus on the shaping of the magnetic field structure due to the structure formation process, 
we ran the simulation without including any additional physics like cooling or star-formation.

\begin{table*}
  \begin{tabular}{rlccr}
  \hline \hline
  & $M_{dm} [M_\odot]$ & $M_{gas} [M_\odot]$ & Grav. Softening $[\kpc]$ & Halo $N_{part}$ \\
  \hline
  1x	& $1.6\times10^9$      & $2.4\times10^8$	   & $7.0$	& $1.5\times10^5$	\\	
  6x	& $2.5\times10^8$      & $3.7\times10^7$	   & $3.6$ 	& $9.7\times10^5$	\\
  10x	& $1.6\times10^8$      & $3.0\times10^7$	   & $3.2$ 	& $1.4\times10^6$	\\
  130x	& $1.2\times10^7$      & $2.3\times10^6$	   & $1.6$ 	& $2.1\times10^7$        \\
  \hline \hline
  \end{tabular}
  \label{tab:0}
  \caption{Overview of the different initial conditions. The last column shows the final total number of particles particles inside the virial radius.}
\end{table*}

\subsection{Galaxy cluster slices and profiles}

\bFIG
  \subfigure[Density (130x)]
  {\includegraphics[width=0.45\textwidth]{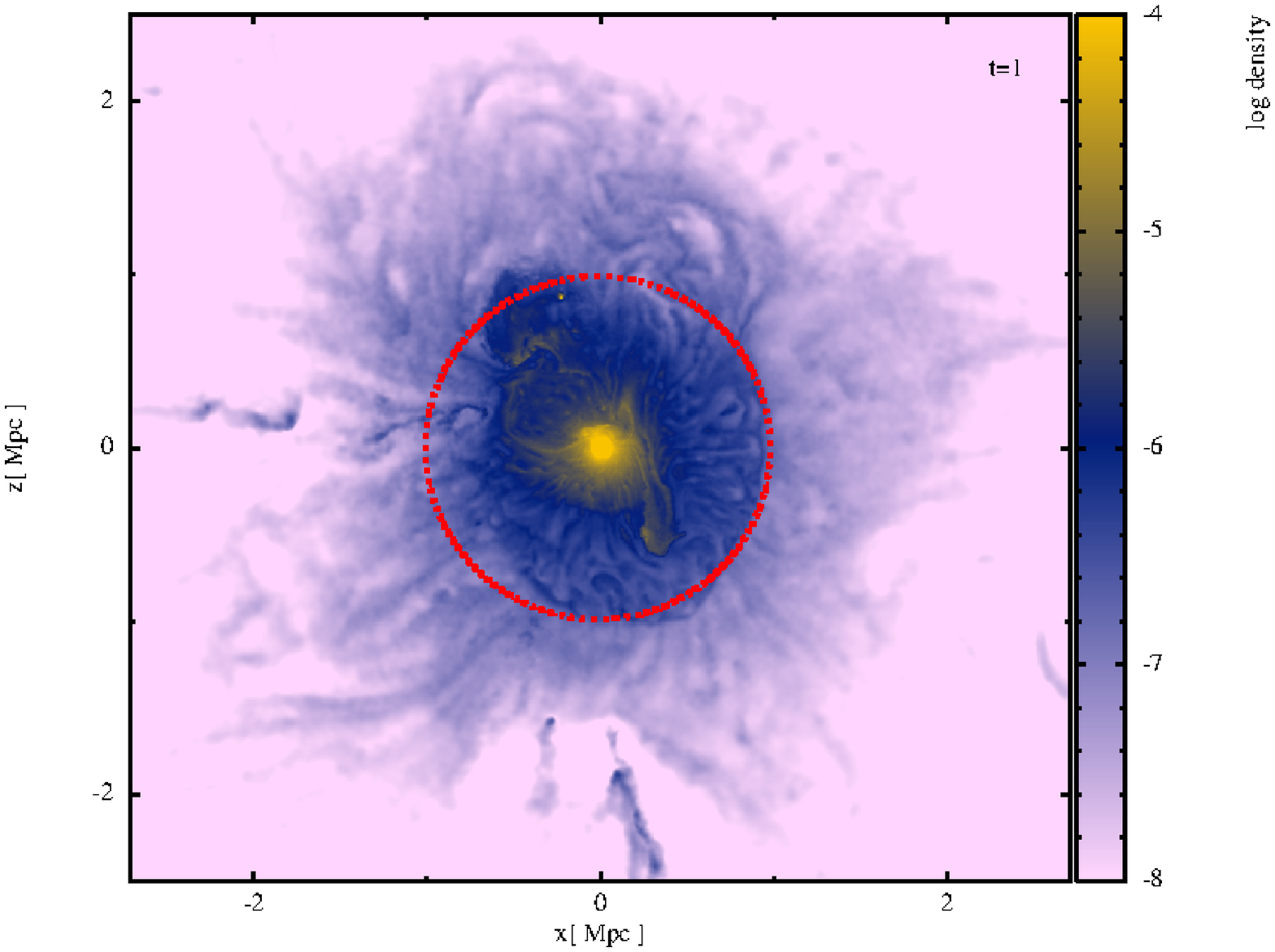}}
  \subfigure[Temperature (130x)]
  {\includegraphics[width=0.45\textwidth]{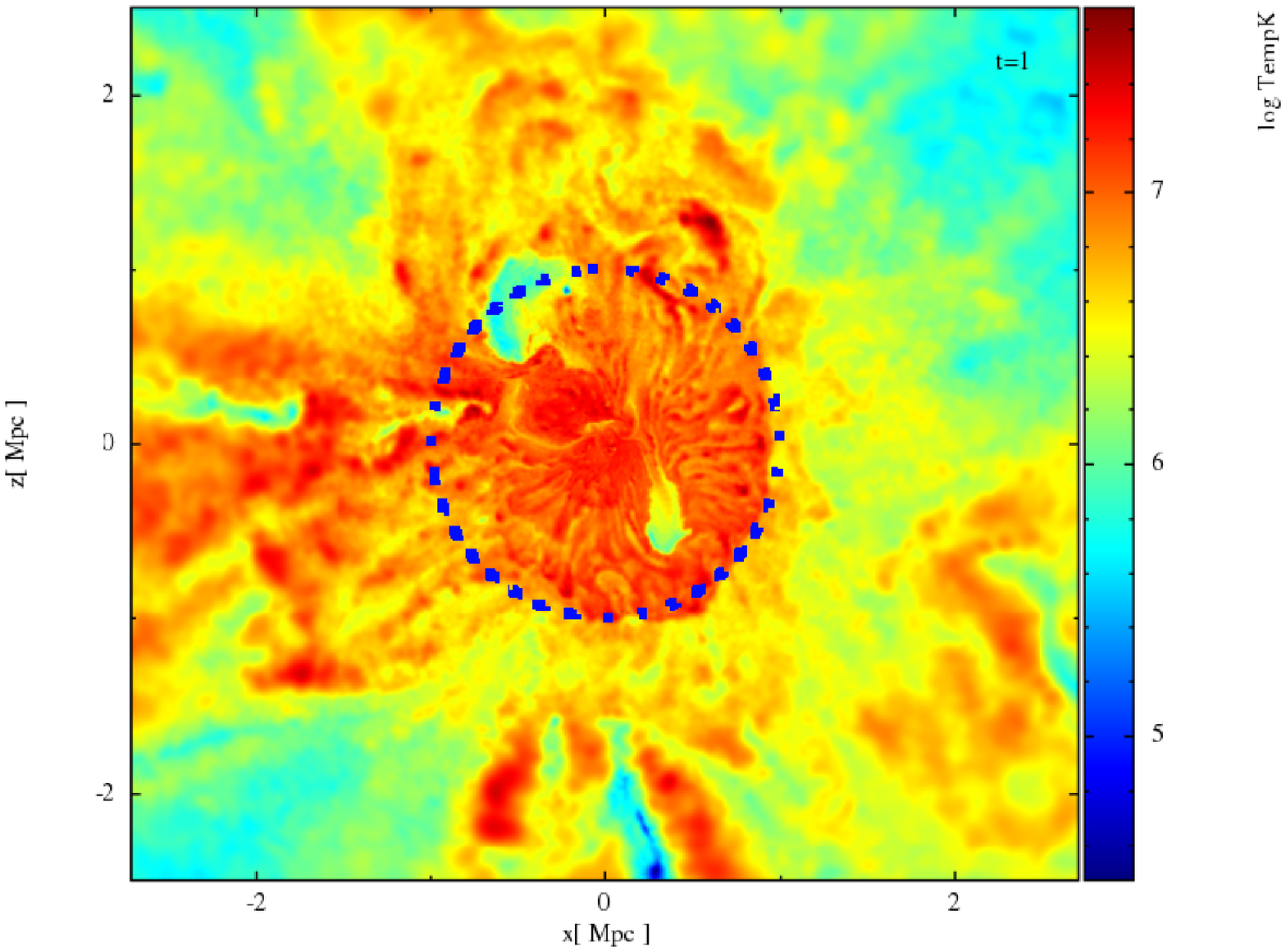}}\\
  \subfigure[Magnetic Energy (130x)]
  {\includegraphics[width=0.45\textwidth]{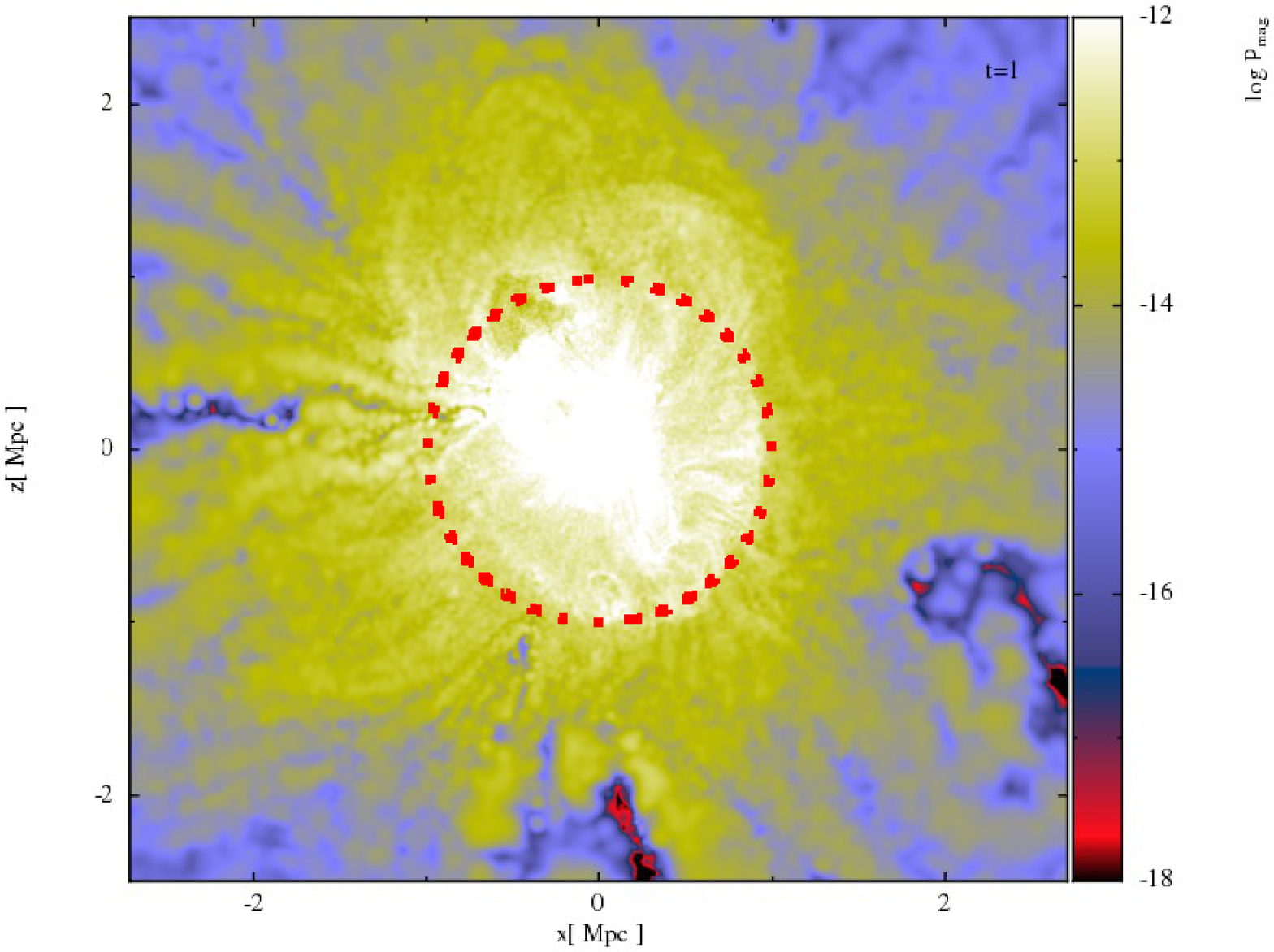}}
  \subfigure[$\divb$ errors (130x)]
  {\includegraphics[width=0.45\textwidth]{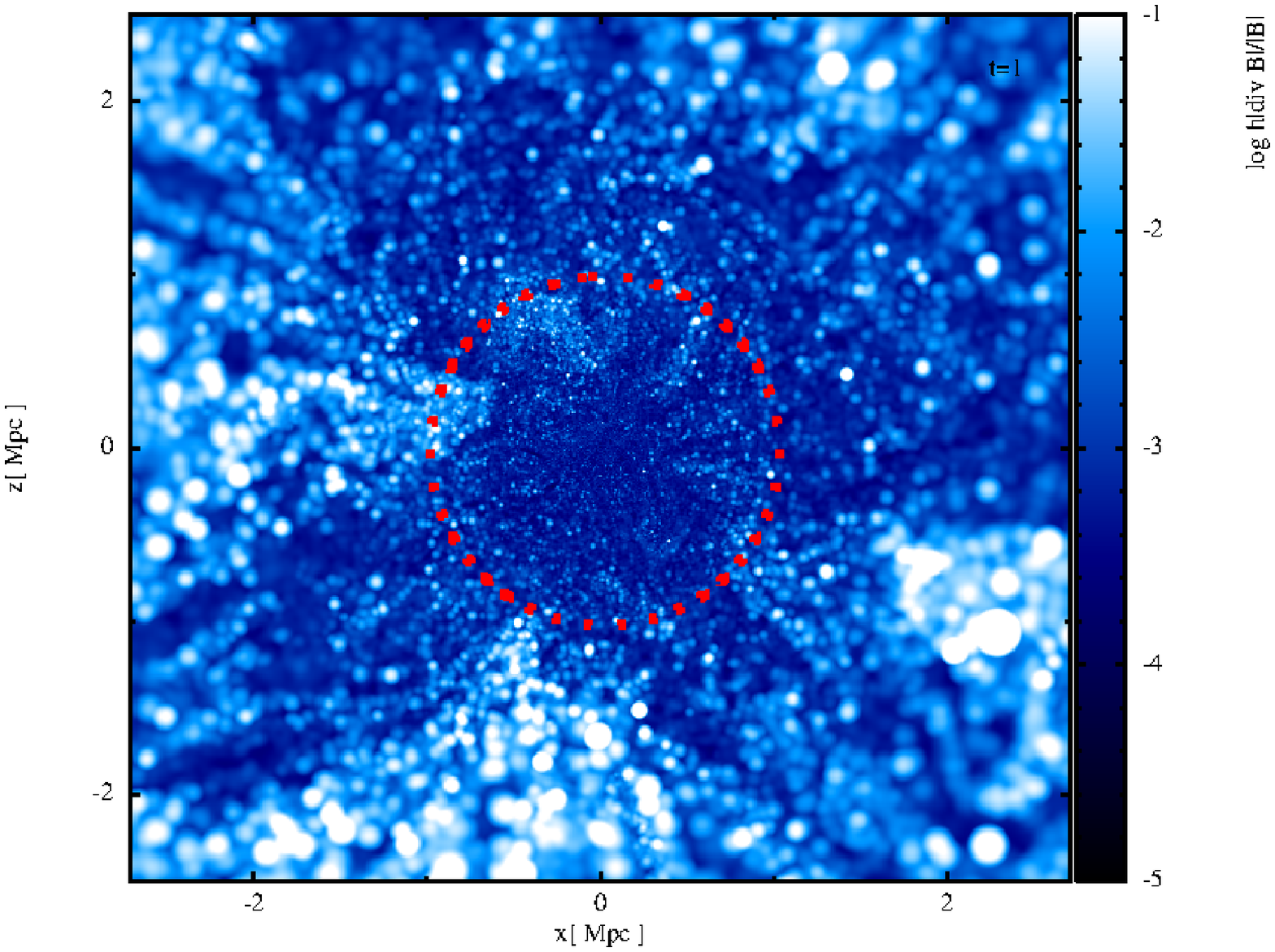}}
  \caption[]{Slices through the center of the galaxy cluster at redshift $z=0$ for 130x resolution, showing the large scale cluster environment. 
  Shown are the density, temperature, magnetic energy and the $\divb$ errors. The dashed circle indicates the virial radius of the cluster.}
\eFIG{Slice}

To show the complexity of the cluster atmosphere we cut thin slices through the center of the highest resolution version of the simulation (130x), 
centered on the potential minimum within the galaxy cluster at $z=0$ in \FIG{Slice}.
Clearly, the accretion shock at the outskirts ($r>2$ Mpc) is visible, getting penetrated by cold, filamentary structures plunging into the hot, cluster atmosphere.
Also several internal shocks (from previous merger events) are still visible within the outer parts of the atmosphere, just approaching the virial radius.
In contrary, the central part of the cluster looks relatively relaxed.
In a high $\beta$ plasma as the simulated here, the magnetic field lines are frozen into the velocity field, while being compressed and advected together.
Therefore, especially beyond the virial radius, the structures seen in the magnetic field are well correlated to the slightly denser gas, 
left over material of several thousands of resolved substructures, where the gas was stripped while the sub-structures got dissolved within 
the denser environment of the cluster during its assembly.
Within the virial radius, structures are more virialized and mixed.
Additionally, filamentary structures within the magnetic field can be noticed due to the presence of turbulence.
Also, additional amplification at the shock fronts, with a mild decline afterwards as expected, are clearly visible.
The distribution of the numerical $\divb$ errors shows that they are mostly located in the very low density region, 
where the resolution of the underlying particle distribution is quite low.

In \FIG{bprof} we show the calculated radial profiles of the magnetic field.
With enough resolution (to resolve the turbulence properly), the magnetic field saturates at roughly equipartition of an fraction of the turbulent energy fraction 
(a fraction of 10 percent would lead to values of order of $10\mu$G).
Naturally from thermodynamics, the energy content of the magnetic field should be comparable to the turbulent energy, which is also observed in galaxies.
This is the result of a turbulent dynamo operating, extensively demonstrated by numerical simulations of colliding galaxies \citep{Kotarba2011,Annette2012} 
and simulations of galactic halo formation \citep{Alex2012}.
In \citet{Alex2012} it was even shown, that the evolution of the magnetic field observed in simulations very well can be described by a simple, turbulent dynamo model.
However, in galaxy clusters the observed magnetic field is found only to be order of $\mu$G, so significantly below equipartition values from the level of turbulence 
present in galaxy clusters.
Consistent with previous studies \citep{Bonafede11} we found that without explicitly modeling physical dissipative processes, 
the magnetic field saturates at larger values than indicated by the observations. 
It is important to stress here, that the numerical $\divb$ errors are decreasing with increasing resolution (\FIG{bprof}), while the magnetic field profile stays converged.
Also, with the exception of the lowest resolution run (1x), the results of the standard \textsc{Spmhd} simulations compare well with the Dedner implementation, regardless of the fact that the Dedner shows less numerical $\divb$ error.
In general, within the 130x simulation, the numerical $\divb$ error is even more than one order of magnitude lower than for the 6x run, 
while leaving the radial magnetic field profile unchanged.
This is another strong prove that the results are not driven by numerical artifacts. 

\bFIG
  \begin{center}
  \subfigure[Magnetic Energy (130x)]
  {\includegraphics[width=0.45\textwidth]{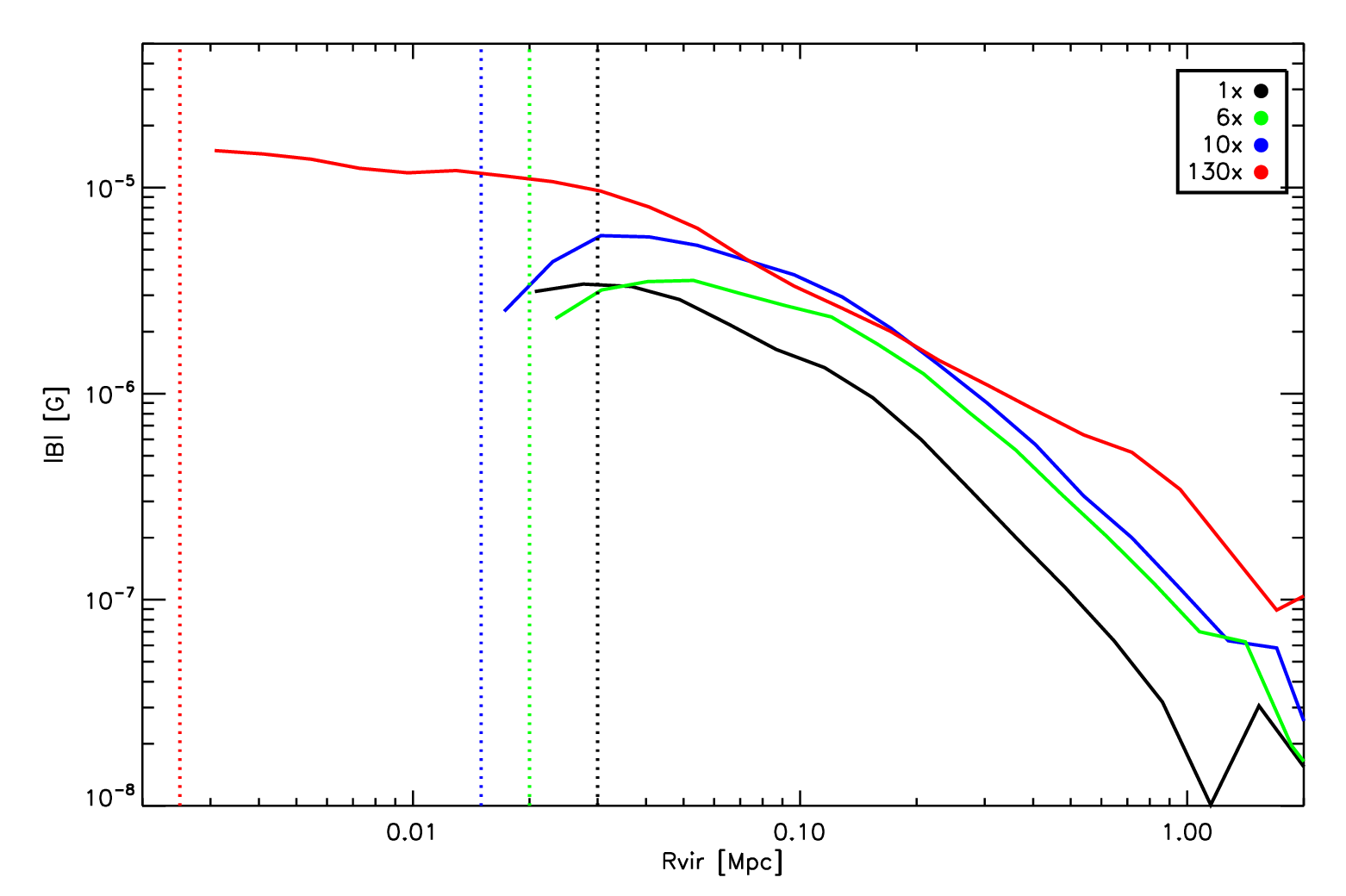}}
  \subfigure[$\divb$ errors (130x)]
  {\includegraphics[width=0.45\textwidth]{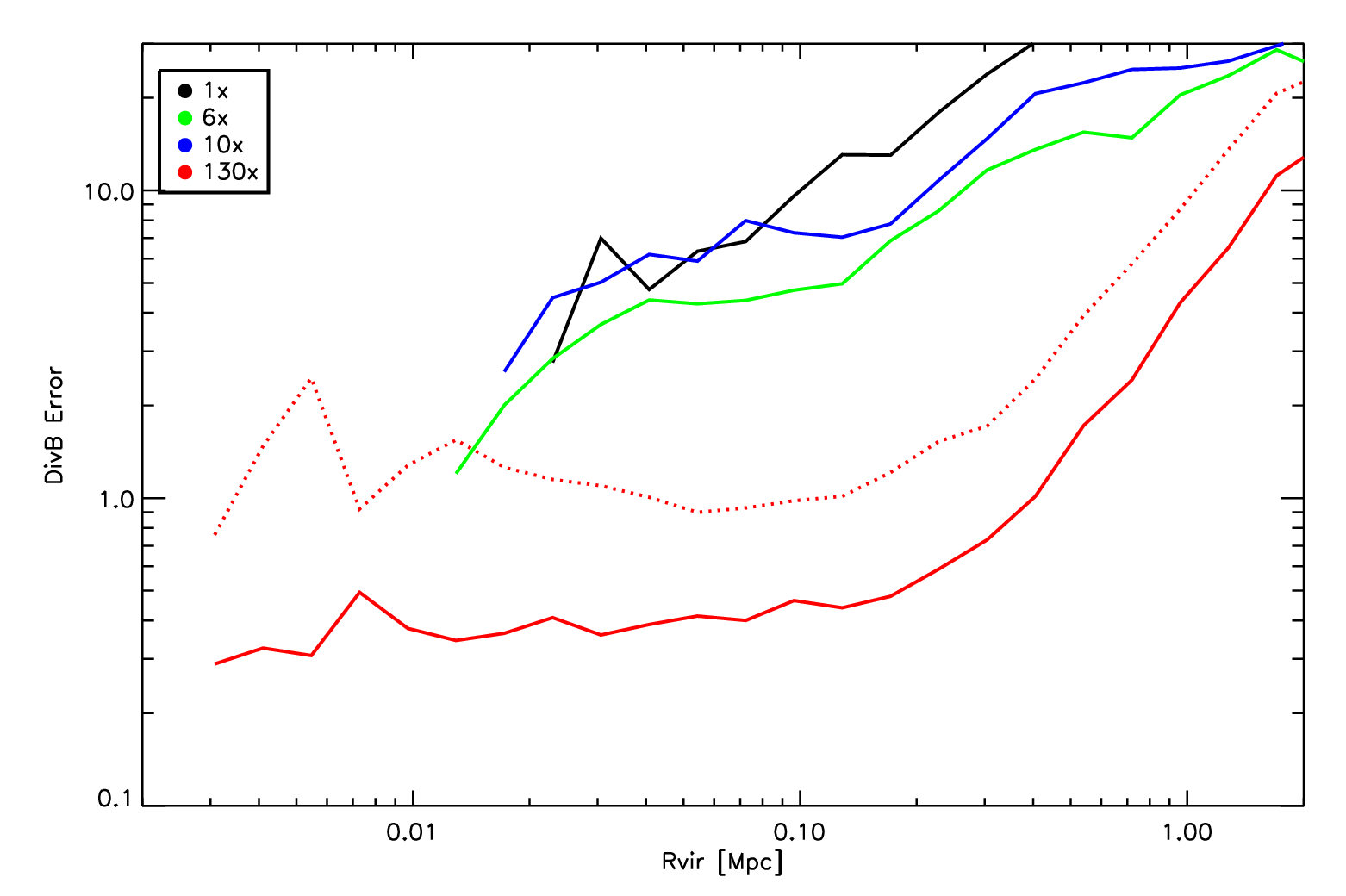}}
  \end{center}
  \caption[]{Radially averaged, magnetic field profiles at $z=0$ for different resolutions (left panel) obtained with the Dedner implementation. 
  The colored lines correspond to different resolutions. Numerical profiles of the $\divb$ errors (right panel). 
  Additionally we show the 130x run without cleaning scheme in dotted lines. Increasing resolution yields lower errors and a better resolved magnetic field.}
\eFIG{bprof}

\subsection{Synthetic RM Measurements} \label{sec:RM}

\bFIG
  \begin{center}
  \subfigure[Standard]
  {\includegraphics[height=0.20\textwidth]{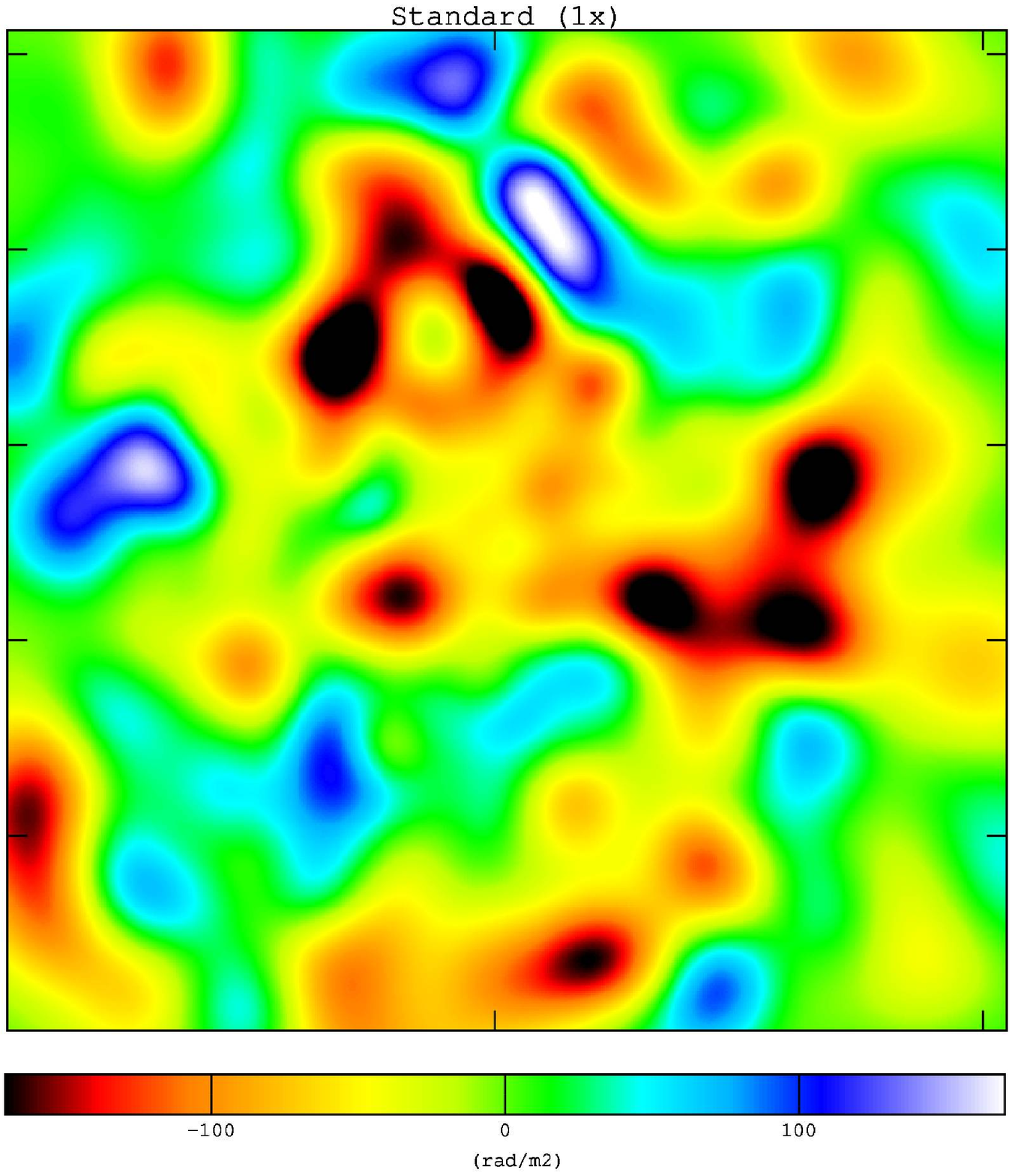}}
  \subfigure[Dedner]
  {\includegraphics[height=0.20\textwidth]{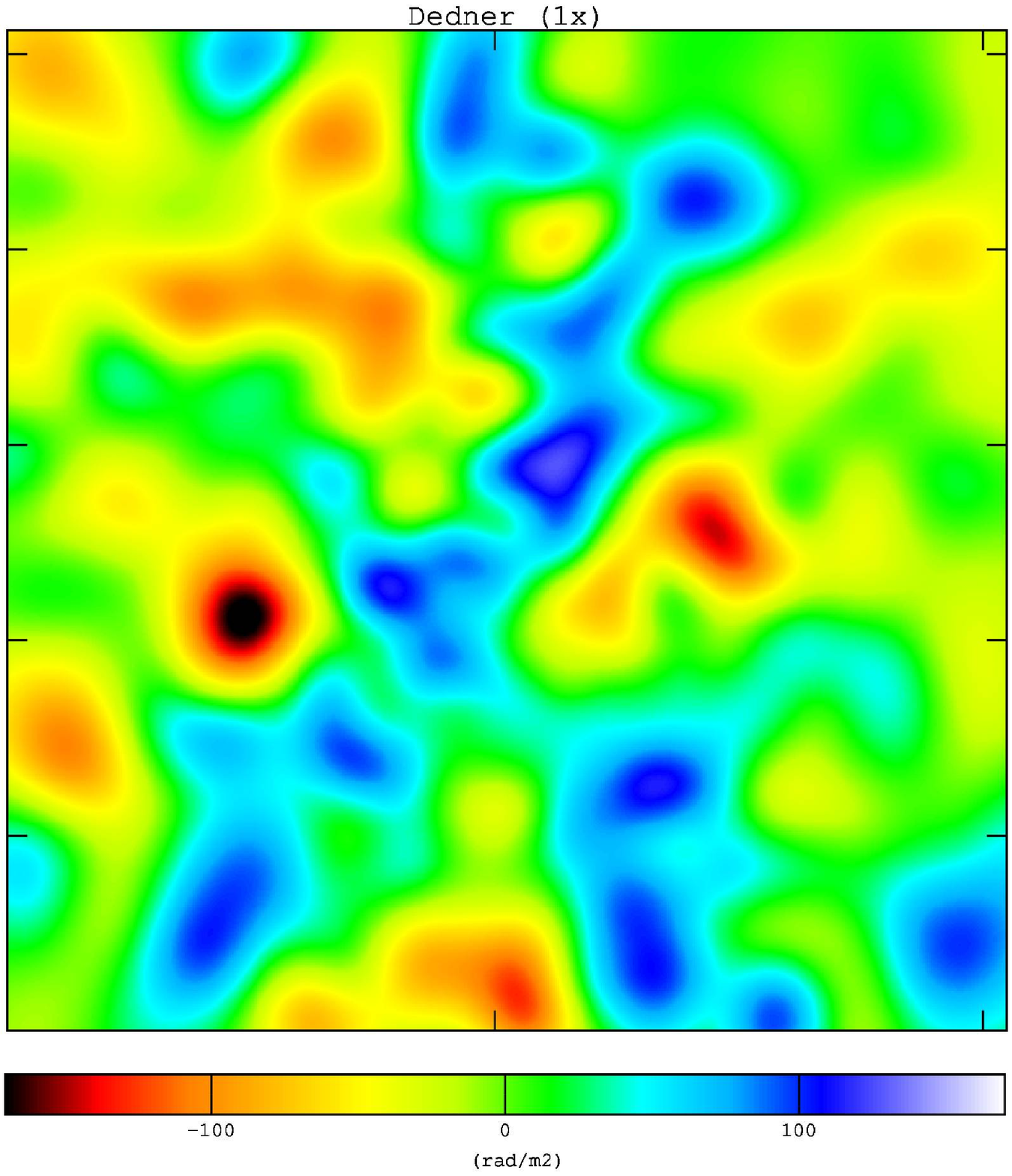}}
  \subfigure[Art. Dis.]
  {\includegraphics[height=0.20\textwidth]{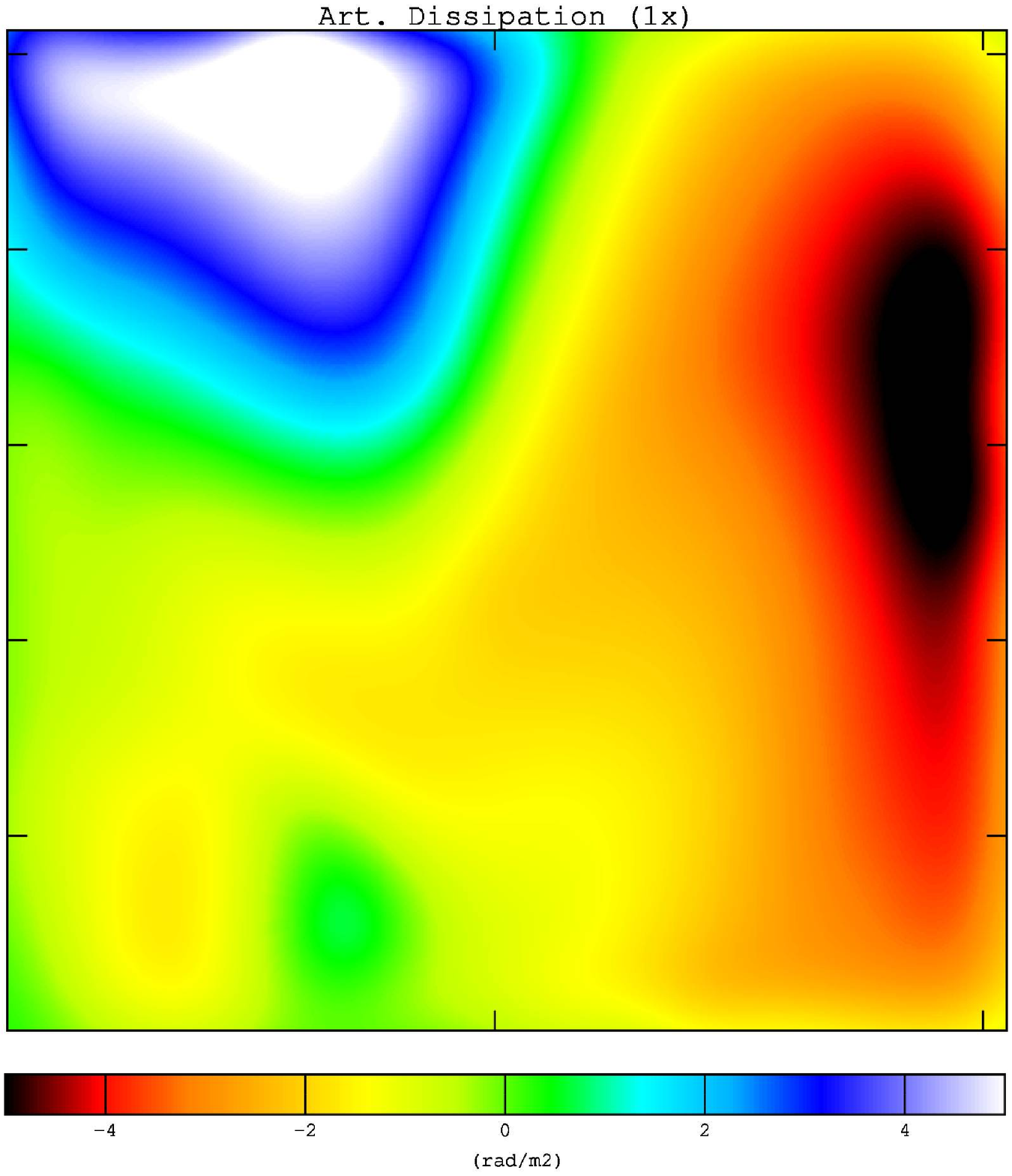}}
  \subfigure[B-Smooth]
  {\includegraphics[height=0.20\textwidth]{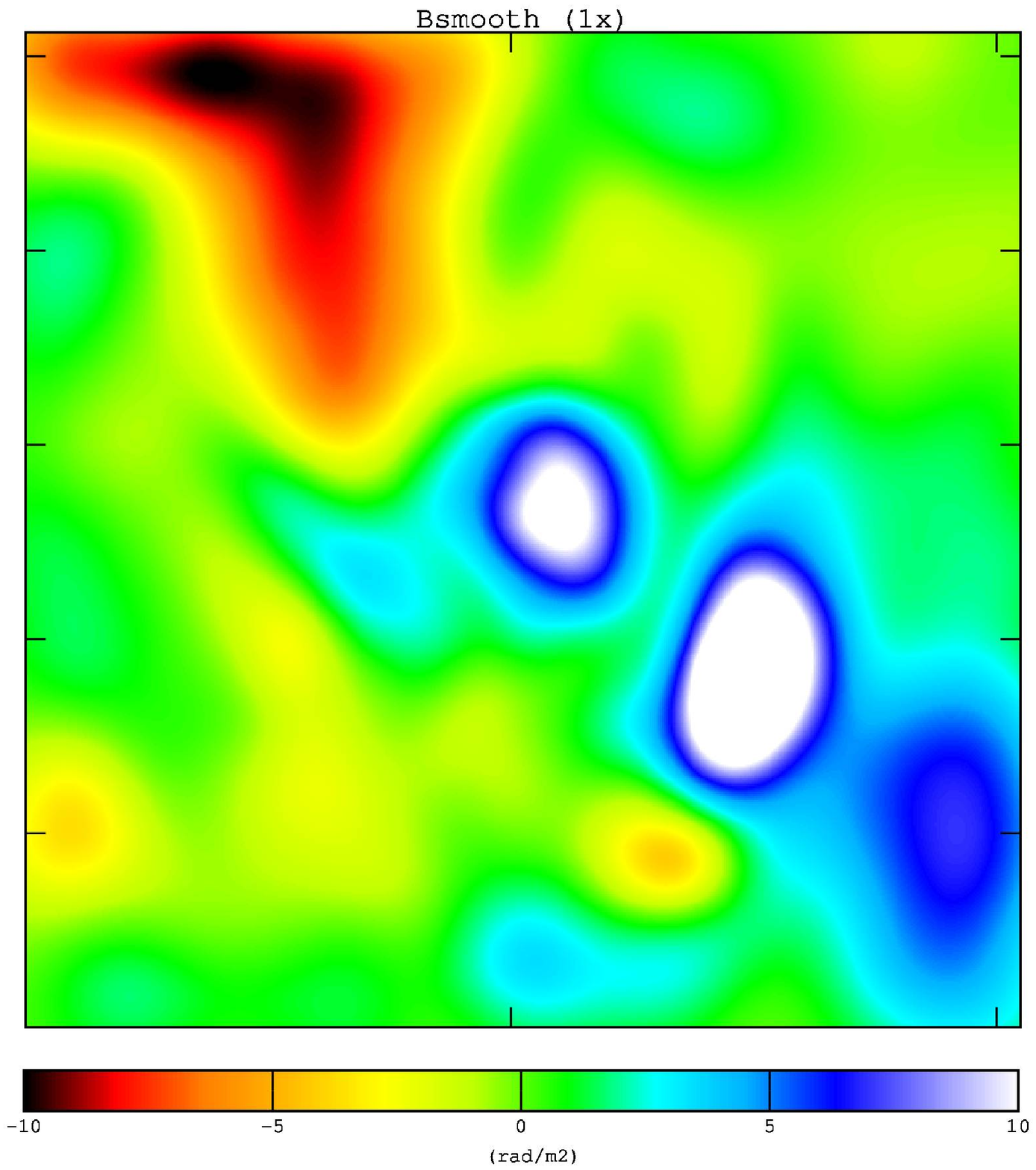}}
  \subfigure[Euler]
  {\includegraphics[height=0.20\textwidth]{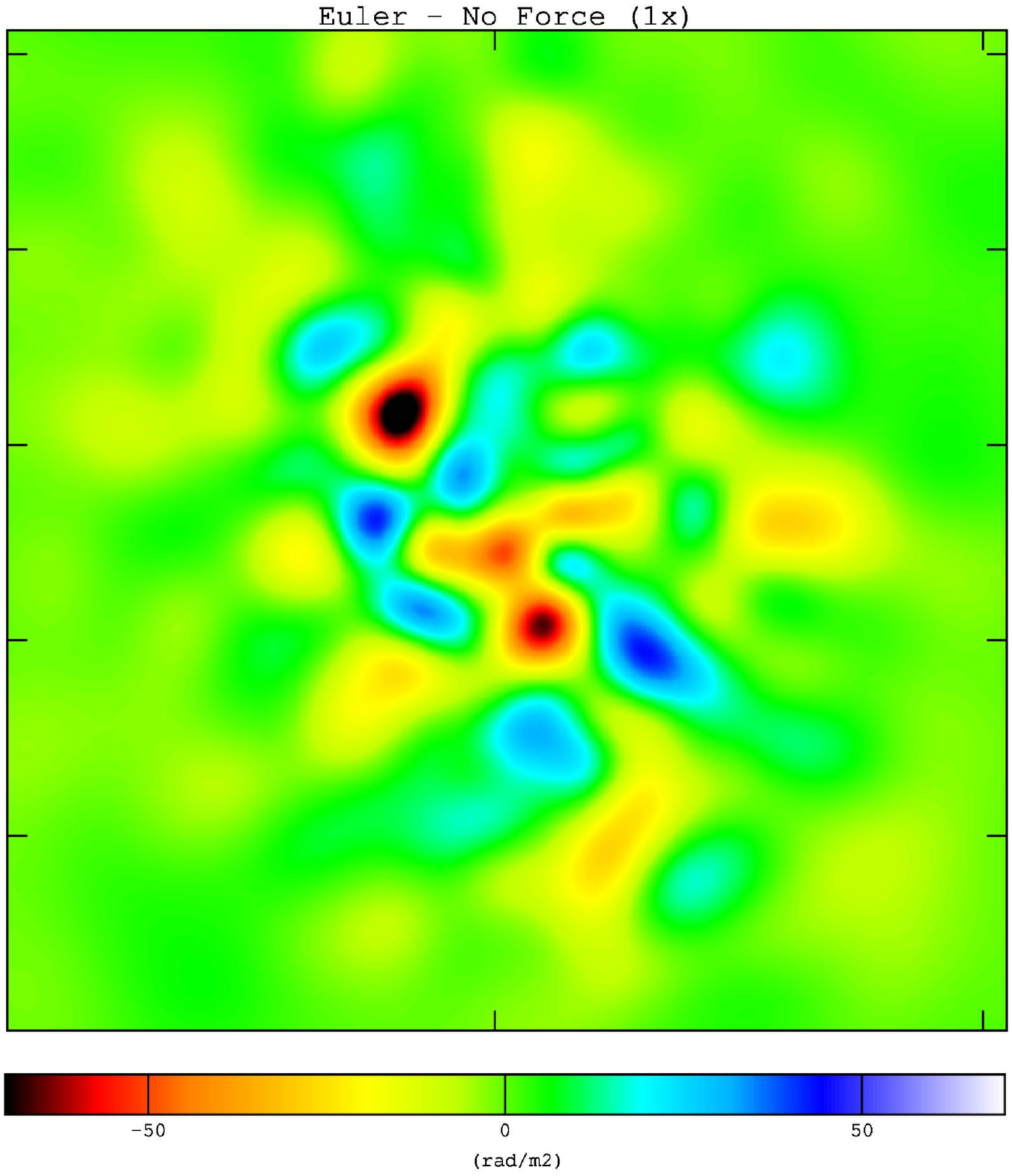}}
  \end{center}
  \caption[]{Synthetic RM maps of the central region of galaxy clusters from the 1x run, using various MHD implementations. 
  From left to right: standard \textsc{Spmhd}, Dedner cleaning, artificial dissipation, B smoothing and Euler potentials without Lorentz force. 
  Shown is a projected region of $174\times174\kpc$ centered onto the cluster.}
\eFIG{rm_schemes}

To compare with observations and especially to study the predicted magnetic field structures within the simulations, we produced synthetic RM maps, 
using \textsc{Smac} \citep{Dolag2005}, which performs a line of sight integration and projection of the simulation SPH data onto a grid.
The simulated cluster was not constrained to reproduce a certain, observed cluster in particular and we in any case neglect the effect 
of a physically motivated magnetic dissipation, which would be needed to match the exact radial magnetic profile of galaxy clusters.
The simulations also neglect the effects of cooling, star-formation and stellar, as well as AGN feedback, which would be needed to reproduce 
certain aspects of the thermal structure within galaxy clusters.
Therefore, we do not expect to match the exact amplitude of the RM signal of the cluster and therefore investigate only the structural properties of the obtained RM maps.

\FIG{rm_schemes} gives an impression of the RM maps obtained from the 1x resolution simulations.
The structural properties of the RM patterns agree very well between the standard implementation and the implementation based on the Dedner scheme.
We here show also the results obtained from the Euler Potential implementation, which passively 
(e.g. neglecting the Lorenz force) evolves the magnetic field in the simulation.
As this implementation just reflects the integrated winding of the magnetic field by the hydrodynamic patterns within the ICM, 
it demonstrates that the magnetic field patterns observed in galaxy clusters is strongly related to the underlying turbulence within the ICM.
We also show the effect of numerical dissipation and magnetic regularization on the predicted magnetic field structures.
it can be seen that numerical magnetic dissipation can lead to severe smoothing of the magnetic field within such simulations.

\bFIGs
  \begin{center}
  \includegraphics[width=0.45\textwidth]{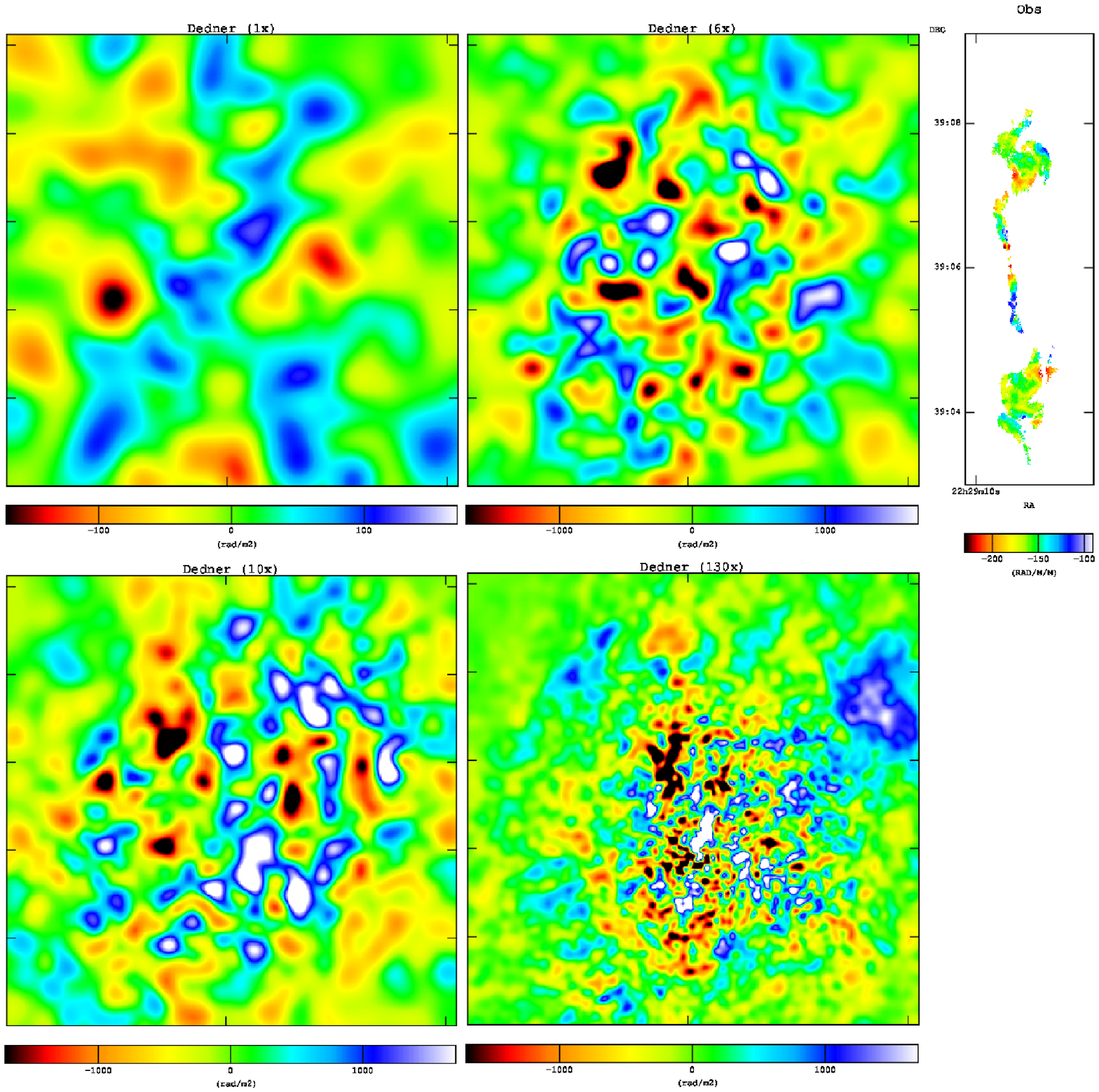}
  \end{center}
  \caption[]{Synthetic RM maps for different resolutions with the Dedner scheme. 
  The maps show the central region ($174\times174~\kpc$) of the simulated cluster. From upper left to lower right: 1x, 6x, 10x and 130x resolution.}
\eFIGs{rm_res}

In \FIG{rm_res} we show synthetic maps obtained with the Dedner implementation at the different resolutions.
With increasing resolution, more magnetic field reversals are resolved.
This is expected as with higher resolution we are able to resolve smaller structures within the turbulent velocity field, which twist and bend the magnetic field.
Note, that with the highest resolution (e.g. the 130x) we reach a resolution, 
which comes close to the coherence lengths of the observed magnetic field on clusters of a few $\kpc$.
For example, the radio lobes of 3C449 \citep{1995A&A...302..680F} can be used to probe the magnetic field of the cluster atmosphere by its RM imprint.
The radio source 3C449 is located in the center of a relatively low mass galaxy cluster with a observed temperature of the ICM 
of $\approx1$kev \citep{1998MNRAS.296.1098H} at $z=0.1711$.
\FIG{rm_comp} compares the observed RM patterns with the RM patterns obtained from the simulations, 
cutting the synthetic maps to the shape of the observed radio lobe, mimicking the observational window.
Both, simulations and observations not just show random patterns but also some more filamentary structures as expected from the presence of MHD turbulence.

\bFIGs
  \begin{center}
  \includegraphics[width=0.35\textwidth]{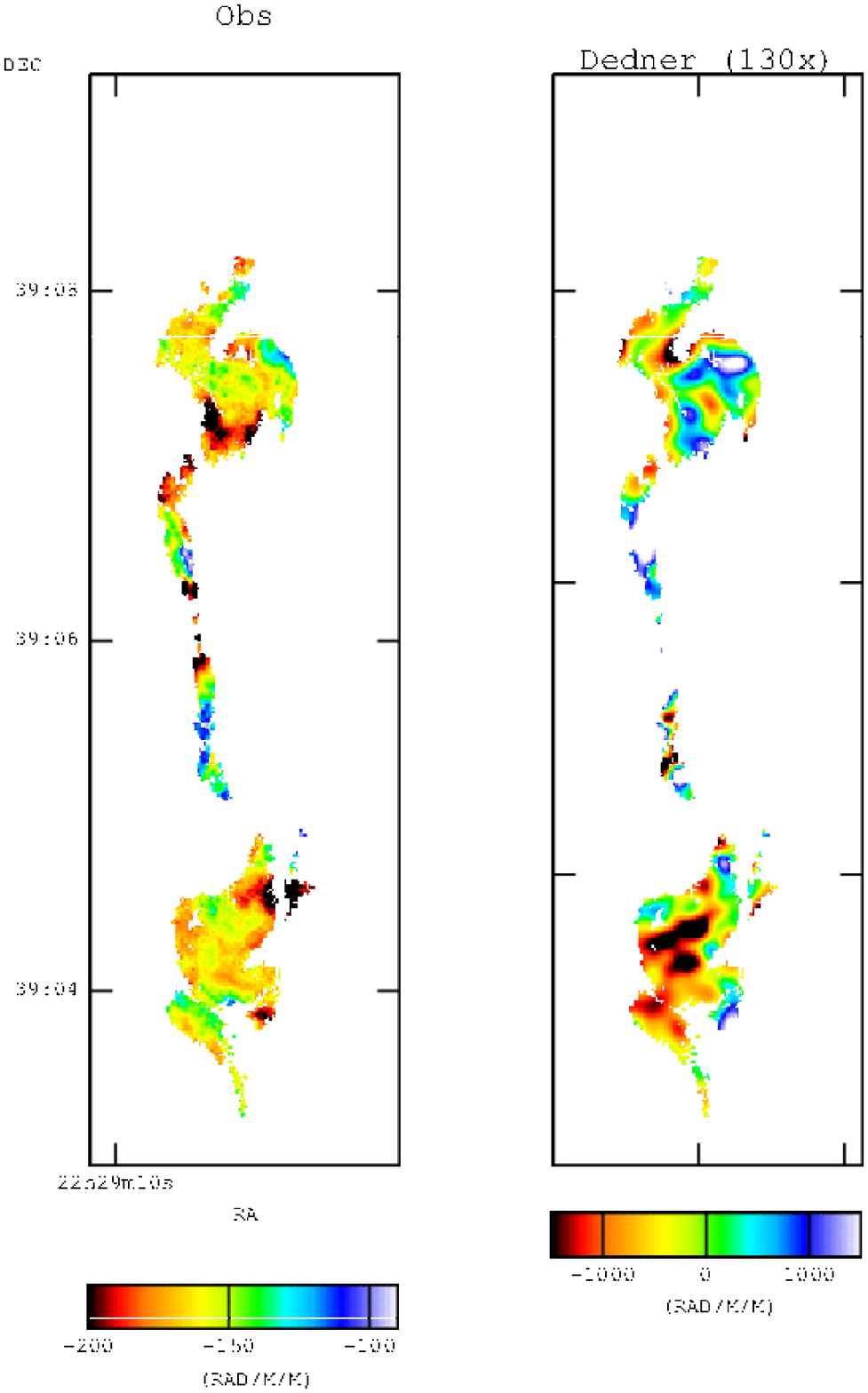}
  \end{center}
  \caption[]{Comparison of the predicted RM map from simulations at the highest resolution (e.g. 130x) with the observed map of 3C449 in \cite{1995A&A...302..680F}. 
  The synthetic map is cut to the same shape as the observational map to allow a better comparison. Both maps have a resolution of approximately $0.17\kpc$ per pixel.}
\eFIGs{rm_comp}

\subsection{Structure Functions}

To take the observational limitations into account when comparing the results of our simulations with observations, 
we used synthetic observations and masked them with the shape of the Faraday screen, as done in \FIG{rm_comp}.
This mimics any kind of bias introduced by the observational window - defined by the size and shape of the observed radio lopes - for our synthetic calculations.
The observed RM maps also show a significant constant offset caused by the galactic magnetic field \cite{1995A&A...302..680F}, 
so we subtract this galactic contribution from the observations when comparing with theoretical predictions.

To obtain a quantitative comparison we calculated the projected second order structure function 
\begin{equation}S(r) = \left<\left( \mathrm{RM}(r') - \mathrm{RM}(r'+r)\right)^2\right>,\end{equation}
of the observed and synthetic RM maps, with $r=\sqrt{\Delta x^2+\Delta y^2}$ being the distance from a pixel at position $r'=(x,y)$.
The resulting matrix is then averaged in radial bins to obtain the structure functions.

\cite{Kraichnan65} formulated one of the first phenomenological theory of MHD turbulence.
He found that in the presence of a strong mean magnetic field, will result in an magnetic energy spectrum of $\sim k^{-3/2}$.
However, if the magnetic field is weak, as used in most of the mean field theories, 
the expected magnetic energy spectrum should be Kolmogorov like and $\sim k^{-5/3}$.
Both theories assume isotropic turbulence, which might not the case in astrophysical systems, 
leading to a suppression of the energy cascade along the direction of the mean magnetic field.
However, observations on galactic scales \citep{Han04} and numerical simulations \citep{Mason08}, 
favor the Kolmogorov spectrum and energy transfer from larger to small scales.

Following \cite{1988ASSL..133.....R}, rough estimates for the structure functions for the RMs can be derived. 
There is a coherence length $l_0$, determining when the RM are not longer specially correlated. 
Thus, for scales $d \ll l_0$
\begin{equation}S(d)\simeq \mathrm{RM}_{0}^{2}\left(\frac{d}{l_0}\right)^{\gamma}\end{equation}
with values of $\gamma=2/3$ for the Kolmogorov spectrum and $\gamma=1/3$ for the Kraichnan spectrum.
At scales larger than $l_0$ the structure functions should have a constant value $\mathrm{RM}_0^2$ and the RMs should be uncorrelated.

\bFIGs
  \includegraphics[width=0.45\textwidth]{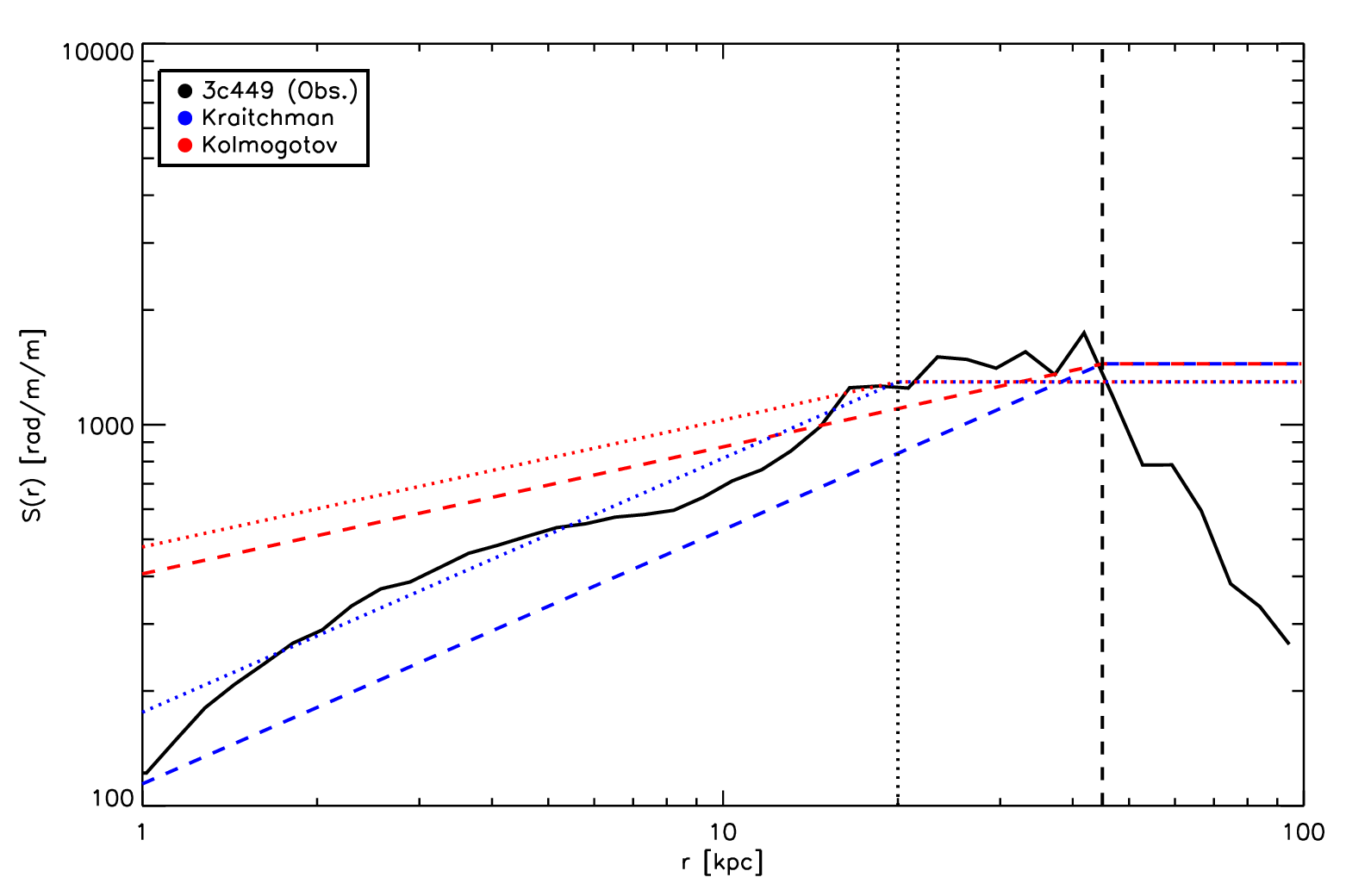}
\caption[Structure functions comparing simple, theoretical expectations with the observations]
{Structure function derived from the observed rotation measurement (black line) of 3C449 in \cite{1995A&A...302..680F} 
compared to the expectation of simple, theoretical models. In red the expected shape for the Kolmogorov spectrum 
and in blue the one expected from a Kraichnan spectrum. The dotted lines correspond the parameter values of $\mathrm{RM}_0=35~\mathrm{rad~m}^{-2}$ 
and $\mathrm{l}_0=20~ \kpc$, whereas the dashed lines correspond to $\mathrm{RM}_0=38~\mathrm{rad~m}^{-2}$ and $\mathrm{l}_0=45\kpc$.
Note, that different values for the parameters allow for different fits of the slopes.}
\eFIGs{RM_theo}

In \FIG{RM_theo} we compare theoretical predictions for the structure functions of those simplified models with the structure 
function computed from the observed RM of 3C449 \citet{1995A&A...302..680F}.
In red is shown the expected shape for a Kolmogorov like spectrum and in blue the expectation for a Kraichnan spectrum.
Unfortunately, the current observations do not allow to distinguish these two models.
The Kolmogorov like spectrum seems to fit slightly better, in line with the findings for the magnetic field power spectrum 
reconstructed from the RMs in Hydra \citep{2011A&A...529A..13K}.
But at larger distances the RM should be uncorrelated and therefore have a constant structure function. 
As the spatial range over which the magnetic field can be measured is still limited to the core region of the cluster, 
density and magnetic field are not significantly declining over the region covered by the radio lobes and therefore can not explain 
the decrease of the structure function at larger distances.
However, the synthetic RM maps and the windowing introduced by the shape of the RM region where  
data is available, can responsible for the radial decline of the structure function at these distances (see \FIGp{Str_final}).
 
\bFIGs
  \includegraphics[width=0.45\textwidth]{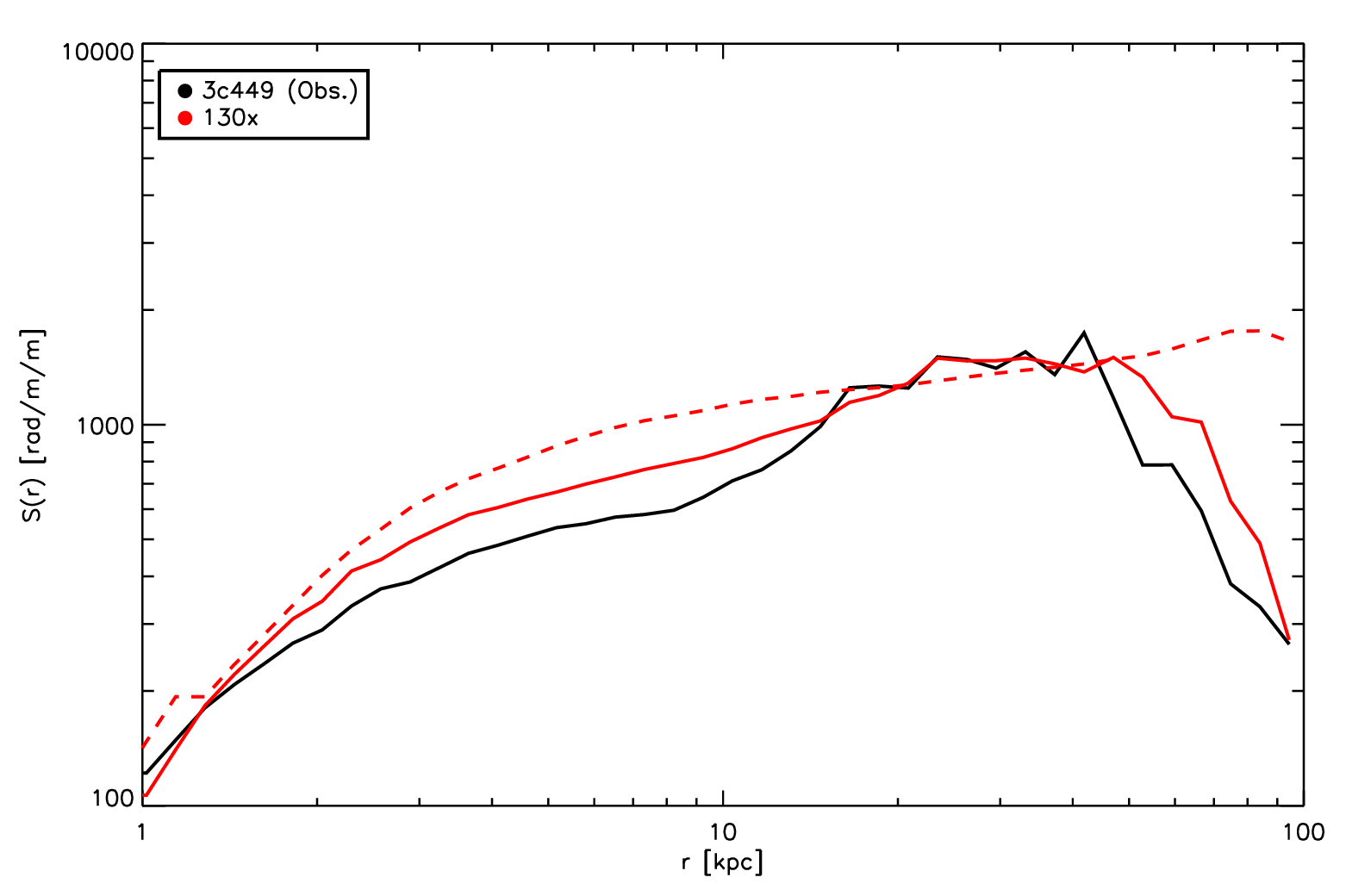}
  \caption[]{Structure functions calculated from the observed RM map (black line) of 3C449 in \cite{1995A&A...302..680F} 
  compared to 130x simulations using the Dedner implementation (red lines). 
  The dashed line shows the structure function calculated from the full ($174\times174$ kpc) maps (see \FIG{rm_res}), 
  whereas the solid lines show maps cut to the shape of the observed RM map (as shown in \FIG{rm_comp}). 
  The dotted lines indicate changes in the slope at around a scale of $10\sim20 \kpc$, visible in the structure function obtained from the simulations.}
\eFIGs{Str_final}

A quantitative comparison between the structural properties of the observed RM and structures obtained from the simulations is shown in \FIG{Str_final}.
As mentioned and discussed earlier, we were interested in the shape of the structure function and therefore re-normalize the amplitude of synthetic RM to the same, 
global RM value of the observations to make them comparable. 
The shape of the structure function at the highest resolution is relatively flat at scales above $10\sim20 \kpc$, 
and shows a steepening at scales smaller than that.
The extreme resolution of our simulation is reflected in the fact, 
that the structure function does not drop to zero even at the smallest scales showing still significant structures.
Especially when the calculation is restricted to the spatial observed mask in 3C449 it matches the observations very well.
How the results change for different diffusive numerical schemes or at different resolution is shown in appendix \ref{App:01}.
As soon as simulations are reaching resolutions similar to the observed scales, 
the magnetic field structure obtained by the simulated galaxy clusters, explains well the magnetic field structures observed through RMs. 
This means that structure formation, by its hierarchical nature and the related turbulence within galaxy clusters is already shaping 
the magnetic field structure within galaxy clusters to observed structures.


\section{Conclusions} \label{sec:conc}

We continue the development started by \citet{DolagStasyszyn09} of an \textsc{Spmhd} implementation of MHD in the cosmological simulation code \textsc{Gadget}. 
We performed various standard test problems and discussed instability corrections, regularization schemes and $\divB$ cleaning techniques. 
The code was applied to cosmological astrophysical simulations, studying the role of resolution and regularization schemes in simulations of galaxy clusters.
The main focus was set on the role of the $\divb = 0$ constrain, also comparing with Eulerian solutions. 
Our main findings can be summarized as follows:

\begin{itemize}
\item 	Correcting the tensile instability by explicitly subtracting the contribution of a numerical non-zero divergence of the magnetic field
	to the Lorenz force from the Maxwell tensor as suggested by \cite{2001ApJ...561...82B} or \cite{DolagStasyszyn09} is performing well. 
	To avoid spurious effects due to sampling problems, particularly in the front shocks, a threshold in the correction improves the performance and quality.
	This feature turns out to be fundamental for simulations with extremely high spatial resolution, where previously 
	the numerical instabilities dominated and the simulations could not have been performed.

\item 	We successfully implemented a multidimensional divergence cleaning method in \textsc{Spmhd}.
 	The \textsc{Gadget} \textsc{Spmhd} implementation continues to perform very well on our multidimensional shock tube tests as well as on commonly used planar test problems. 
	We showed that the Dedner cleaning scheme does not affect the shape of shocks, but reduces the $\divb$ errors. 
	This is important in astrophysical situations. 
	Also, the evolution of the cleaning is completely local and only affected by the \textsc{Spmhd} interpolants.

\item 	Testing our schemes in long time runs of the Orszag-Tang Vortex, we found that regularization schemes which depend on dissipation as well as implementations based on
	Euler potentials can lead to instabilities, whereas the standard implementation as well as the Dedner cleaning scheme are found to be robust and stable.
	This is a warning on the limits of possible MHD implementations within astrophysical environments. 
\end{itemize}

We obtained our most important findings by comparing with an Euler (divergence-free by construction) scheme. 
This shows, that we have already reached the $\divb$ error levels inherent from the \textsc{Spmhd} scheme itself 
(see \FIGp{VortexDivbe}), and any further cleaning or regularization will require to numerically dissipate the magnetic field. 
However, this does not yield an error-free implementation, thus still some $\divb$ related instabilities can arise.
In the case of the Euler potentials, we stress the fact that this representation lacks important features in the evolution of the magnetic field \citep{2010MNRAS.401..347B}.
However, outstanding progress has been made on providing stable MHD and \textsc{Spmhd} implementations, removing unwanted instabilities.

Furthermore, our code has different features with each regularization or cleaning scheme used.
The success of the Dedner cleaning brings the simulations closer to the ``ideal'' MHD state, 
by locally calculating and subtracting the error term in the induction equation.

In \FIG{comp_imp} we show a comparison between all the implementations in different tests. 
The dissipative schemes have lower $\divb$ errors, however the Dedner cleaning schemes stands out, 
lowering the $\divb$ error, and can in most cases improve the results. 
In contrast, there are some tests showing that to lowering $\divb$ errors can over-smooth important features in contrast to a correct solution.

We ran cosmological simulations of galaxy clusters with different underlying \textsc{Spmhd} schemes for the magnetic field evolution. 
The magnetic field profiles of the different simulation schemes show similar properties, even when comparing with Eulerian codes.

The comparison between the different schemes leads us to the conclusion that in general, 
the non-vanishing numerical $\divb$ terms are not important for the proper evolution or growth of the magnetic field. 

It is known, that in galaxy clusters, the growth of the magnetic field is not only driven by the adiabatic collapse. 
The magnetic fields lines are strongly tangled on all scales, leading to a complex, turbulent morphology \citep{1988ASSL..133.....R}. 
An intrinsic feature of \textsc{Spmhd} is its adaptivity, particularly in spatial resolution. 
Therefore, the turbulent cascade is well described also within the framework of the collapse of cosmological structures, 
leading to a physical growth of the magnetic field by a turbulent dynamo.
Note, that previous work clearly demonstrated that equipartition between the magnetic energy and the turbulent energy is reached, 
independent of the magnetic seed field strength. 
Therefore, the predicted magnetic field structure within galaxy clusters can be studied independently of the real origin of the magnetic seed fields. 
By analyzing synthetic RM maps, we evaluated that previously used dissipative regularization schemes are producing numerically driven, 
large reversal scales within the magnetic field, which are not comparable to what is seen in observations. 
However, our standard \textsc{Spmhd} implementation as well as our Dedner implementation allow to follow the magnetic field
structures without loosing resolution by dissipative effects. 
In the extremely high resolution simulations we performed, magnetic field structures down to $\kpc$ scales are for the first time resolved
within cosmological simulations of structure formation.
By studying structure functions of synthetic RM maps, we found that these simulations reproduce very well the structures within
observed RM maps. This demonstrates that the observed structure of the magnetic field within galaxy clusters can be shaped already
by the cosmological structure formation process and the thereby driven, turbulent gas motions.

Finally, the use of advanced \textsc{Spmhd} schemes allows to simulate the properties of the ICM at very high spatial resolutions, 
contributing to the global picture of the properties of the hot gas within galaxy clusters.


\section*{acknowledgements}

Rendered plots were made using \textsc{Splash} \citep{2007PASA...24..159P}.
KD is supported by the DFG Priority Programme 1177 and by the DFG 
Cluster of Excellence \textquoteleft Origin and Structure of the Universe\textquoteright.
FAS is supported by the DFG Research Unit 1254.
FAS thanks the useful discussions with Axel Brandeburg, Daniel Price and Florian B\"urzle, during the development of this project.
\bibliographystyle{mn2e}
\bibliography{master}

\begin{thebibliography}{}

\bibitem[\protect\citeauthoryear{{Anderson}, {Hirschmann}, {Liebling} \&
  {Neilsen}}{{Anderson} et~al.}{2006}]{Anderson2006}
{Anderson} M.,  {Hirschmann} E.~W.,  {Liebling} S.~L.,    {Neilsen} D.,  2006,
  Classical and Quantum Gravity, 23, 6503

\bibitem[\protect\citeauthoryear{{Balsara} \& {Spicer}}{{Balsara} \&
  {Spicer}}{1999}]{Balsara99}
{Balsara} D.~S.,  {Spicer} D.~S.,  1999, Journal of Computational Physics, 149,
  270

\bibitem[\protect\citeauthoryear{{Beck}, {Lesch}, {Dolag}, {Kotarba}, {Geng} \&
  {Stasyszyn}}{{Beck} et~al.}{2012}]{Alex2012}
{Beck} A.~M.,  {Lesch} H.,  {Dolag} K.,  {Kotarba} H.,  {Geng} A.,
  {Stasyszyn} F.~A.,  2012, \mnras, p.~2729

\bibitem[\protect\citeauthoryear{{Beck}}{{Beck}}{2009}]{2009ASTRA...5...43B}
{Beck} R.,  2009, Astrophysics and Space Sciences Transactions, 5, 43

\bibitem[\protect\citeauthoryear{{Bonafede}, {Dolag}, {Stasyszyn}, {Murante} \&
  {Borgani}}{{Bonafede} et~al.}{2011}]{Bonafede11}
{Bonafede} A.,  {Dolag} K.,  {Stasyszyn} F.,  {Murante} G.,    {Borgani} S.,
  2011, \mnras, 418, 2234

\bibitem[\protect\citeauthoryear{{Bonafede}, {Feretti}, {Murgia}, {Govoni},
  {Giovannini}, {Dallacasa}, {Dolag} \& {Taylor}}{{Bonafede}
  et~al.}{2010}]{Bonafede10}
{Bonafede} A.,  {Feretti} L.,  {Murgia} M.,  {Govoni} F.,  {Giovannini} G.,
  {Dallacasa} D.,  {Dolag} K.,    {Taylor} G.~B.,  2010, \aap, 513, A30+

\bibitem[\protect\citeauthoryear{{B{\o}rve}, {Omang} \& {Trulsen}}{{B{\o}rve}
  et~al.}{2001}]{2001ApJ...561...82B}
{B{\o}rve} S.,  {Omang} M.,    {Trulsen} J.,  2001, \apj, 561, 82

\bibitem[\protect\citeauthoryear{{B{\o}rve}, {Omang} \& {Trulsen}}{{B{\o}rve}
  et~al.}{2006}]{2006ApJ...652.1306B}
{B{\o}rve} S.,  {Omang} M.,    {Trulsen} J.,  2006, \apj, 652, 1306

\bibitem[\protect\citeauthoryear{{Brandenburg}}{{Brandenburg}}{2010}]{2010MNRAS.401..347B}
{Brandenburg} A.,  2010, \mnras, 401, 347

\bibitem[\protect\citeauthoryear{{Brio} \& {Wu}}{{Brio} \&
  {Wu}}{1988}]{BrioWu88}
{Brio} M.,  {Wu} C.~C.,  1988, Journal of Computational Physics, 75, 400

\bibitem[\protect\citeauthoryear{{Br{\"u}ggen}, {Ruszkowski}, {Simionescu},
  {Hoeft} \& {Dalla Vecchia}}{{Br{\"u}ggen} et~al.}{2005}]{2005ApJ...631L..21B}
{Br{\"u}ggen} M.,  {Ruszkowski} M.,  {Simionescu} A.,  {Hoeft} M.,    {Dalla
  Vecchia} C.,  2005, \apjl, 631, L21

\bibitem[\protect\citeauthoryear{{Bryan} \& {Norman}}{{Bryan} \&
  {Norman}}{1998}]{bryan1998}
{Bryan} G.~L.,  {Norman} M.~L.,  1998, \apj, 495, 80

\bibitem[\protect\citeauthoryear{{C{\'e}cere}, {Lehner} \&
  {Reula}}{{C{\'e}cere} et~al.}{2008}]{Cecere2008}
{C{\'e}cere} M.,  {Lehner} L.,    {Reula} O.,  2008, Computer Physics
  Communications, 179, 545

\bibitem[\protect\citeauthoryear{{Collins}, {Xu}, {Norman}, {Li} \&
  {Li}}{{Collins} et~al.}{2010}]{2010ApJS..186..308C}
{Collins} D.~C.,  {Xu} H.,  {Norman} M.~L.,  {Li} H.,    {Li} S.,  2010, \apjs,
  186, 308

\bibitem[\protect\citeauthoryear{{Dai} \& {Woodward}}{{Dai} \&
  {Woodward}}{1994}]{1994JCoPh.115..485D}
{Dai} W.,  {Woodward} P.~R.,  1994, Journal of Computational Physics, 115, 485

\bibitem[\protect\citeauthoryear{{Dedner}, {Kemm}, {Kr{\"o}ner}, {Munz},
  {Schnitzer} \& {Wesenberg}}{{Dedner} et~al.}{2002}]{Dedner02}
{Dedner} A.,  {Kemm} F.,  {Kr{\"o}ner} D.,  {Munz} C.-D.,  {Schnitzer} T.,
  {Wesenberg} M.,  2002, Journal of Computational Physics, 175, 645

\bibitem[\protect\citeauthoryear{{Dolag}, {Bartelmann} \& {Lesch}}{{Dolag}
  et~al.}{1999}]{Dolag99}
{Dolag} K.,  {Bartelmann} M.,    {Lesch} H.,  1999, \aap, 348, 351

\bibitem[\protect\citeauthoryear{{Dolag}, {Bartelmann} \& {Lesch}}{{Dolag}
  et~al.}{2002}]{Dolag02}
{Dolag} K.,  {Bartelmann} M.,    {Lesch} H.,  2002, \aap, 387, 383

\bibitem[\protect\citeauthoryear{{Dolag}, {Grasso}, {Springel} \&
  {Tkachev}}{{Dolag} et~al.}{2005}]{dolag05}
{Dolag} K.,  {Grasso} D.,  {Springel} V.,    {Tkachev} I.,  2005, Journal of
  Cosmology and Astro-Particle Physics, 1, 9

\bibitem[\protect\citeauthoryear{{Dolag}, {Hansen}, {Roncarelli} \&
  {Moscardini}}{{Dolag} et~al.}{2005}]{Dolag2005}
{Dolag} K.,  {Hansen} F.~K.,  {Roncarelli} M.,    {Moscardini} L.,  2005,
  \mnras, 363, 29

\bibitem[\protect\citeauthoryear{{Dolag}, {Kachelriess}, {Ostapchenko} \&
  {Tom{\`a}s}}{{Dolag} et~al.}{2011}]{2011ApJ...727L...4D}
{Dolag} K.,  {Kachelriess} M.,  {Ostapchenko} S.,    {Tom{\`a}s} R.,  2011,
  \apjl, 727, L4

\bibitem[\protect\citeauthoryear{{Dolag} \& {Stasyszyn}}{{Dolag} \&
  {Stasyszyn}}{2009}]{DolagStasyszyn09}
{Dolag} K.,  {Stasyszyn} F.,  2009, \mnras, 398, 1678

\bibitem[\protect\citeauthoryear{{Dubois} \& {Teyssier}}{{Dubois} \&
  {Teyssier}}{2008}]{DuboisTeyssier08}
{Dubois} Y.,  {Teyssier} R.,  2008, \aap, 482, L13

\bibitem[\protect\citeauthoryear{{Feretti}, {Dallacasa}, {Giovannini} \&
  {Tagliani}}{{Feretti} et~al.}{1995}]{1995A&A...302..680F}
{Feretti} L.,  {Dallacasa} D.,  {Giovannini} G.,    {Tagliani} A.,  1995, \aap,
  302, 680

\bibitem[\protect\citeauthoryear{{Geng}, {Kotarba}, {B{\"u}rzle}, {Dolag},
  {Stasyszyn}, {Beck} \& {Nielaba}}{{Geng} et~al.}{2012}]{Annette2012}
{Geng} A.,  {Kotarba} H.,  {B{\"u}rzle} F.,  {Dolag} K.,  {Stasyszyn} F.,
  {Beck} A.,    {Nielaba} P.,  2012, \mnras, 419, 3571

\bibitem[\protect\citeauthoryear{{Govoni}}{{Govoni}}{2006}]{2006AN....327..539G}
{Govoni} F.,  2006, Astronomische Nachrichten, 327, 539

\bibitem[\protect\citeauthoryear{{Govoni}, {Dolag}, {Murgia}, {Feretti},
  {Schindler}, {Giovannini}, {Boschin}, {Vacca} \& {Bonafede}}{{Govoni}
  et~al.}{2010}]{2010A&A...522A.105G}
{Govoni} F.,  {Dolag} K.,  {Murgia} M.,  {Feretti} L.,  {Schindler} S.,
  {Giovannini} G.,  {Boschin} W.,  {Vacca} V.,    {Bonafede} A.,  2010, \aap,
  522, A105

\bibitem[\protect\citeauthoryear{{Guidetti}, {Laing}, {Murgia}, {Govoni},
  {Gregorini} \& {Parma}}{{Guidetti} et~al.}{2010}]{Guidetti10}
{Guidetti} D.,  {Laing} R.~A.,  {Murgia} M.,  {Govoni} F.,  {Gregorini} L.,
  {Parma} P.,  2010, \aap, 514, A50

\bibitem[\protect\citeauthoryear{{Han}, {Ferriere} \& {Manchester}}{{Han}
  et~al.}{2004}]{Han04}
{Han} J.~L.,  {Ferriere} K.,    {Manchester} R.~N.,  2004, \apj, 610, 820

\bibitem[\protect\citeauthoryear{{Hardcastle}, {Worrall} \&
  {Birkinshaw}}{{Hardcastle} et~al.}{1998}]{1998MNRAS.296.1098H}
{Hardcastle} M.~J.,  {Worrall} D.~M.,    {Birkinshaw} M.,  1998, \mnras, 296,
  1098

\bibitem[\protect\citeauthoryear{{Iapichino} \& {Niemeyer}}{{Iapichino} \&
  {Niemeyer}}{2008}]{2008MNRAS.388.1089I}
{Iapichino} L.,  {Niemeyer} J.~C.,  2008, \mnras, 388, 1089

\bibitem[\protect\citeauthoryear{{Inogamov} \& {Sunyaev}}{{Inogamov} \&
  {Sunyaev}}{2003}]{2003AstL...29..791I}
{Inogamov} N.~A.,  {Sunyaev} R.~A.,  2003, Astronomy Letters, 29, 791

\bibitem[\protect\citeauthoryear{{Katz} \& {White}}{{Katz} \&
  {White}}{1993}]{1993ApJ...412..455K}
{Katz} N.,  {White} S.~D.~M.,  1993, \apj, 412, 455

\bibitem[\protect\citeauthoryear{{Keppens}, {Meliani}, {van Marle}, {Delmont},
  {Vlasis} \& {van der Holst}}{{Keppens} et~al.}{2012}]{2012JCoPh.231..718K}
{Keppens} R.,  {Meliani} Z.,  {van Marle} A.~J.,  {Delmont} P.,  {Vlasis} A.,
   {van der Holst} B.,  2012, Journal of Computational Physics, 231, 718

\bibitem[\protect\citeauthoryear{{Kotarba}, {Lesch}, {Dolag}, {Naab},
  {Johansson}, {Donnert} \& {Stasyszyn}}{{Kotarba} et~al.}{2011}]{Kotarba2011}
{Kotarba} H.,  {Lesch} H.,  {Dolag} K.,  {Naab} T.,  {Johansson} P.~H.,
  {Donnert} J.,    {Stasyszyn} F.~A.,  2011, \mnras, 415, 3189

\bibitem[\protect\citeauthoryear{{Kraichnan}}{{Kraichnan}}{1965}]{Kraichnan65}
{Kraichnan} R.~H.,  1965, Physics of Fluids, 8, 1385

\bibitem[\protect\citeauthoryear{{Kuchar} \& {En{\ss}lin}}{{Kuchar} \&
  {En{\ss}lin}}{2011}]{2011A&A...529A..13K}
{Kuchar} P.,  {En{\ss}lin} T.~A.,  2011, \aap, 529, A13

\bibitem[\protect\citeauthoryear{{Londrillo} \& {Del Zanna}}{{Londrillo} \&
  {Del Zanna}}{2000}]{LondrilloDelZanna00}
{Londrillo} P.,  {Del Zanna} L.,  2000, \apj, 530, 508

\bibitem[\protect\citeauthoryear{{Mason}, {Cattaneo} \& {Boldyrev}}{{Mason}
  et~al.}{2008}]{Mason08}
{Mason} J.,  {Cattaneo} F.,    {Boldyrev} S.,  2008, Physics Reviews, 77,
  036403

\bibitem[\protect\citeauthoryear{{Miniati} \& {Martin}}{{Miniati} \&
  {Martin}}{2011}]{2011ApJS..195....5M}
{Miniati} F.,  {Martin} D.~F.,  2011, \apjs, 195, 5

\bibitem[\protect\citeauthoryear{{Orszag} \& {Tang}}{{Orszag} \&
  {Tang}}{1979}]{Orzang79}
{Orszag} S.~A.,  {Tang} C.,  1979, Journal of Fluid Mechanics, 90, 129

\bibitem[\protect\citeauthoryear{{Pakmor}, {Bauer} \& {Springel}}{{Pakmor}
  et~al.}{2011}]{Pakmor2011}
{Pakmor} R.,  {Bauer} A.,    {Springel} V.,  2011, \mnras, 418, 1392

\bibitem[\protect\citeauthoryear{{Paul}, {Iapichino}, {Miniati}, {Bagchi} \&
  {Mannheim}}{{Paul} et~al.}{2011}]{2011ApJ...726...17P}
{Paul} S.,  {Iapichino} L.,  {Miniati} F.,  {Bagchi} J.,    {Mannheim} K.,
  2011, \apj, 726, 17

\bibitem[\protect\citeauthoryear{{Picone} \& {Dahlburg}}{{Picone} \&
  {Dahlburg}}{1991}]{1991PhFlB...3...29P}
{Picone} J.~M.,  {Dahlburg} R.~B.,  1991, Physics of Fluids B, 3, 29

\bibitem[\protect\citeauthoryear{{Price}}{{Price}}{2007}]{2007PASA...24..159P}
{Price} D.~J.,  2007, Publications of the Astronomical Society of Australia,
  24, 159

\bibitem[\protect\citeauthoryear{{Price}}{{Price}}{2012}]{Price2010}
{Price} D.~J.,  2012, Journal of Computational Physics, 231, 759

\bibitem[\protect\citeauthoryear{{Price} \& {Monaghan}}{{Price} \&
  {Monaghan}}{2004}]{PriceI}
{Price} D.~J.,  {Monaghan} J.~J.,  2004, \mnras, 348, 123

\bibitem[\protect\citeauthoryear{{Price} \& {Monaghan}}{{Price} \&
  {Monaghan}}{2005}]{PriceIII}
{Price} D.~J.,  {Monaghan} J.~J.,  2005, \mnras, 364, 384

\bibitem[\protect\citeauthoryear{{Rasia}, {Tormen} \& {Moscardini}}{{Rasia}
  et~al.}{2004}]{rasia2004}
{Rasia} E.,  {Tormen} G.,    {Moscardini} L.,  2004, \mnras, 351, 237

\bibitem[\protect\citeauthoryear{{Rosswog} \& {Price}}{{Rosswog} \&
  {Price}}{2007}]{Rosswog07}
{Rosswog} S.,  {Price} D.,  2007, \mnras, 379, 915

\bibitem[\protect\citeauthoryear{{{Ruzmaikin}, A. A. and {Sokolov}, D. D. and
  {Shukurov} A.M.}}{{{Ruzmaikin}, A. A. and {Sokolov}, D. D. and {Shukurov}
  A.M.}}{1988}]{1988ASSL..133.....R}
{{Ruzmaikin}, A. A. and {Sokolov}, D. D. and {Shukurov} A.M.} ed. 1988,
  {Magnetic fields of galaxies} Vol.~133 of Astrophysics and Space Science
  Library

\bibitem[\protect\citeauthoryear{{Ryu} \& {Jones}}{{Ryu} \&
  {Jones}}{1995}]{1995ApJ...442..228R}
{Ryu} D.,  {Jones} T.~W.,  1995, \apj, 442, 228

\bibitem[\protect\citeauthoryear{{Springel}}{{Springel}}{2005}]{springel05}
{Springel} V.,  2005, \mnras, 364, 1105

\bibitem[\protect\citeauthoryear{{Springel}}{{Springel}}{2010}]{Springel2010}
{Springel} V.,  2010, \mnras, 401, 791

\bibitem[\protect\citeauthoryear{{Springel}, {Yoshida} \& {White}}{{Springel}
  et~al.}{2001}]{springel01}
{Springel} V.,  {Yoshida} N.,    {White} S.~D.~M.,  2001, New Astronomy, 6, 79

\bibitem[\protect\citeauthoryear{{Stone}, {Gardiner}, {Teuben}, {Hawley} \&
  {Simon}}{{Stone} et~al.}{2008}]{Stone2008}
{Stone} J.~M.,  {Gardiner} T.~A.,  {Teuben} P.,  {Hawley} J.~F.,    {Simon}
  J.~B.,  2008, \apjs, 178, 137

\bibitem[\protect\citeauthoryear{{Subramanian}, {Shukurov} \&
  {Haugen}}{{Subramanian} et~al.}{2006}]{Subramanian2006}
{Subramanian} K.,  {Shukurov} A.,    {Haugen} N.~E.~L.,  2006, \mnras, 366,
  1437

\bibitem[\protect\citeauthoryear{{Tormen}, {Bouchet} \& {White}}{{Tormen}
  et~al.}{1997}]{tormen97}
{Tormen} G.,  {Bouchet} F.~R.,    {White} S.~D.~M.,  1997, MNRAS, 286, 865

\bibitem[\protect\citeauthoryear{{Toth}}{{Toth}}{2000}]{Toth00}
{Toth} G.,  2000, Journal of Computational Physics, 161, 605

\bibitem[\protect\citeauthoryear{{Tribble}}{{Tribble}}{1991}]{1991MNRAS.253..147T}
{Tribble} P.~C.,  1991, \mnras, 253, 147

\bibitem[\protect\citeauthoryear{{Vacca}, {Murgia}, {Govoni}, {Feretti},
  {Giovannini}, {Perley} \& {Taylor}}{{Vacca} et~al.}{2012}]{Vacca12}
{Vacca} V.,  {Murgia} M.,  {Govoni} F.,  {Feretti} L.,  {Giovannini} G.,
  {Perley} R.~A.,    {Taylor} G.~B.,  2012, \aap, 540, A38

\bibitem[\protect\citeauthoryear{{Vazza}, {Brunetti}, {Gheller}, {Brunino} \&
  {Br{\"u}ggen}}{{Vazza} et~al.}{2011}]{2011A&A...529A..17V}
{Vazza} F.,  {Brunetti} G.,  {Gheller} C.,  {Brunino} R.,    {Br{\"u}ggen} M.,
  2011, \aap, 529, A17

\bibitem[\protect\citeauthoryear{{Vazza}, {Brunetti}, {Kritsuk}, {Wagner},
  {Gheller} \& {Norman}}{{Vazza} et~al.}{2009}]{2009A&A...504...33V}
{Vazza} F.,  {Brunetti} G.,  {Kritsuk} A.,  {Wagner} R.,  {Gheller} C.,
  {Norman} M.,  2009, \aap, 504, 33

\bibitem[\protect\citeauthoryear{{Vazza}, {Tormen}, {Cassano}, {Brunetti} \&
  {Dolag}}{{Vazza} et~al.}{2006}]{2006MNRAS.369L..14V}
{Vazza} F.,  {Tormen} G.,  {Cassano} R.,  {Brunetti} G.,    {Dolag} K.,  2006,
  \mnras, 369, L14

\bibitem[\protect\citeauthoryear{{White}}{{White}}{1996}]{1996clss.conf..349W}
{White} S.~D.~M.,  1996, in {Schaeffer} R.,  {Silk} J.,  {Spiro} M.,
  {Zinn-Justin} J.,  eds, Cosmology and Large Scale Structure {Formation and
  Evolution of Galaxies}.
pp 349--+

\bibitem[\protect\citeauthoryear{{Xu}, {Li}, {Collins}, {Li} \& {Norman}}{{Xu}
  et~al.}{2009}]{Xu2009}
{Xu} H.,  {Li} H.,  {Collins} D.~C.,  {Li} S.,    {Norman} M.~L.,  2009, \apjl,
  698, L14

\bibitem[\protect\citeauthoryear{{Xu}, {Li}, {Collins}, {Li} \& {Norman}}{{Xu}
  et~al.}{2010}]{2010ApJ...725.2152X}
{Xu} H.,  {Li} H.,  {Collins} D.~C.,  {Li} S.,    {Norman} M.~L.,  2010, \apj,
  725, 2152

\end{thebibliography}

\begin{appendix}

\section{Dependence of the structure function on the numerical setup}
\label{App:01}

Comparing the structure functions obtained from simulations using different
numerical implementations and underlying resolutions, demonstrates 
how effective these simulations can describe the detailed
magnetic field structure observed in the RM maps of galaxy clusters.

\bFIGs
  \includegraphics[width=0.45\textwidth]{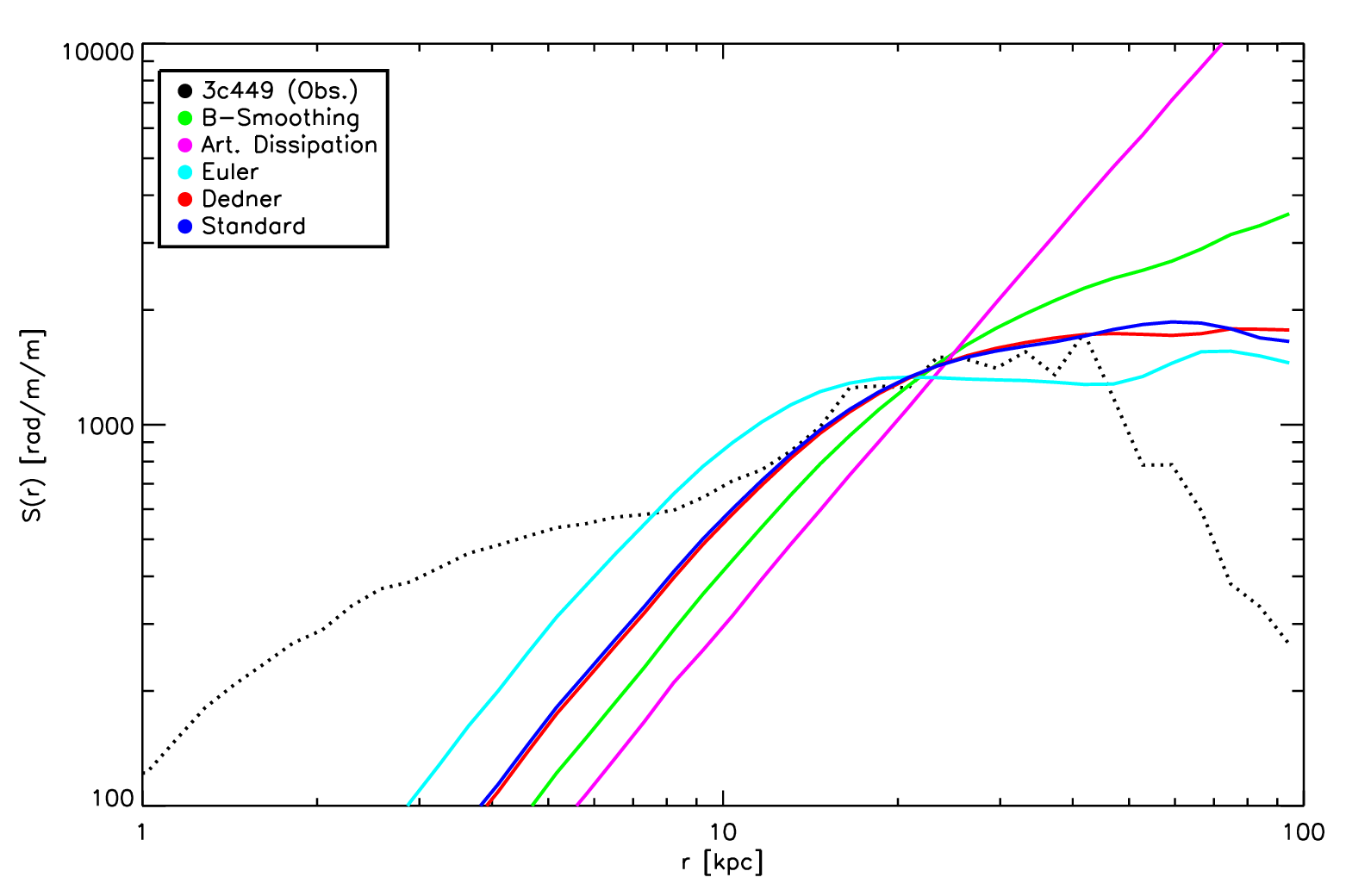} 
\caption[Structure function in different resolution.]
{Structure function $S$ calculated from full ($174\times174~\kpc$) maps (see \FIGp{rm_res}) for different implementations, but all at 1x resolution.
As a reference we plot the result of the function calculated from the observed rotation measurement (doted black line) of 3C449 in \cite{1995A&A...302..680F}.
}
\eFIGs{Str_show_full_sch}

In the case of comparing different implementations, we can study the effective resolution reached
due to the numerical dissipation within the different implementations. 

For each simulated cluster we calculated the synthetic RM maps with a physical size of $174\times174~\kpc$ 
(see \FIGp{rm_schemes}) and calculate their structure function $S$, shown in \FIG{Str_show_full_sch}, 
ignoring the windowing which would be introduced by the visibility of the real observations.
To account for the varying efficiency to resolve the underlaying dynamo mechanism among the different numerical implementations 
and to make them comparable with the observations, we rescaled the synthetic maps using the as a reference the amplitude at large scales
in observations, assuming that at those scales the RMs are uncorrelated. 
For the schemes, where the regularization is based on dissipation,
the effective resolution is drastically reduced by large factors compared to the SPH smoothing scale.
Therefore the resolved coherence length at the 1x resolution is reduced to scales comparable to the
size of the maps used and manifests itself in a structure function which still keeps growing.

\bFIGs
  \includegraphics[width=0.45\textwidth]{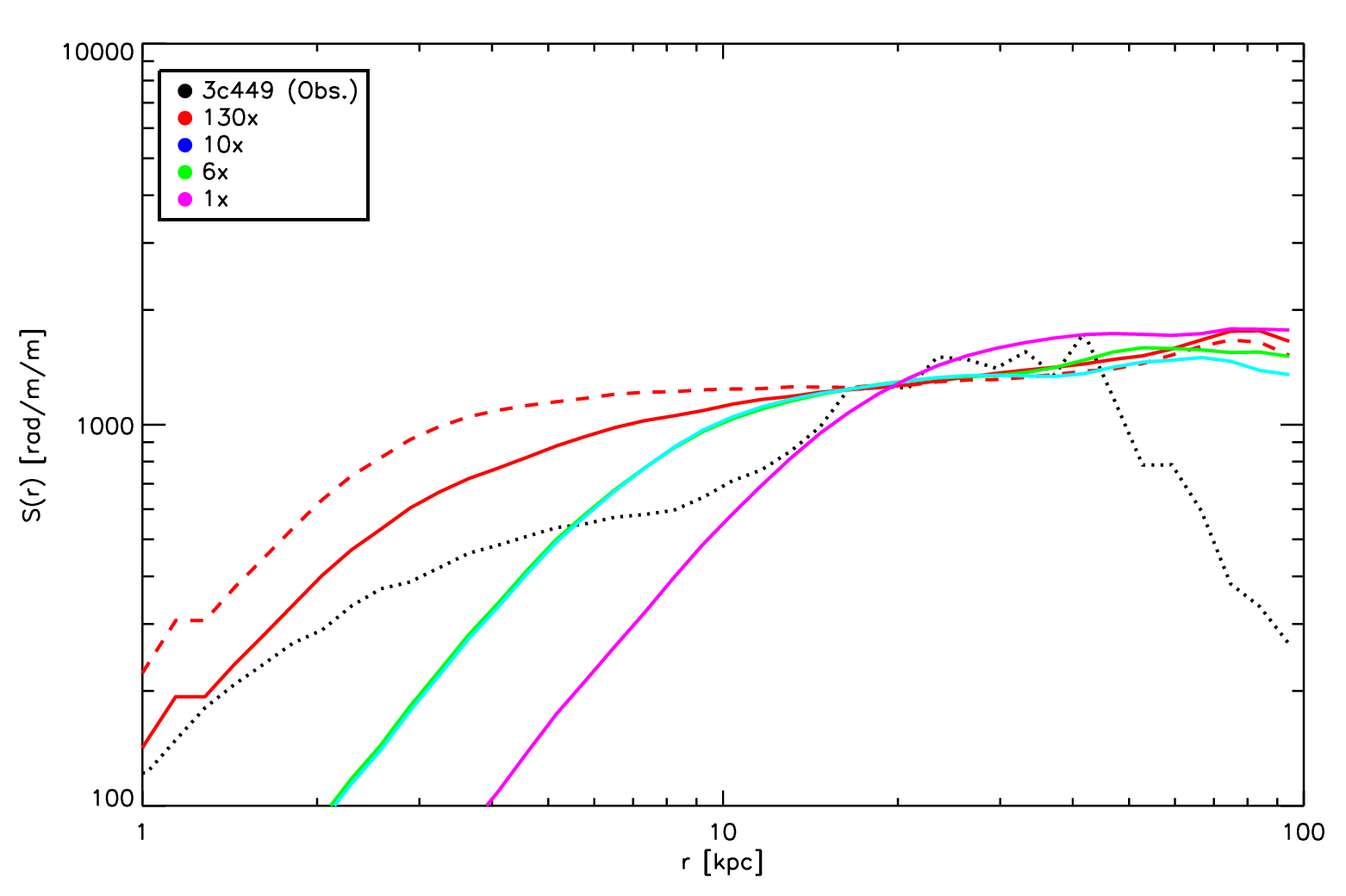} 
\caption[Structure function in different resolution.]
{Shown are structure functions $S$ calculated from the full ($174\times174~\kpc$) maps (see \FIGp{rm_res}) 
for the simulations with increasing resolution using the Dedner implementation and in dashed the highest 
resolution for the standard case (i.e. without cleaning scheme).
As a reference we show the result of the from the observed rotation measurement (doted black line).
}
\eFIGs{Str_show_res}

In \FIG{Str_show_res} a resolution study using the Dedner implementation is shown by the structure functions 
calculated from the $174\times174~\kpc$  RM maps. 
As before, we scaled the synthetic maps using the amplitude of the observations.
Increasing the resolution we are able to resolve smaller scales in the underlying turbulence, reflected in being able
to reach smaller and smaller scales within the RM maps. Also, the coherence length of the turbulent field shifts to smaller and
smaller scales, as visible by the beginning of the flat part moving towards smaller scales when increasing resolution.
There are no considerable differences between the 6x and 10x runs,  as the increase in resolution is quite minor in that case.
This demonstrates the need to vary the resolution by a large amount when investigating resolution effects, as done here by 
increasing the resolution overall by a factor of 130.
In the 130x case, we also show the case without Dedner cleaning, where the power at small scales is even larger 
and the coherence scale is moved very close to the resolution scale. Here it can not be excluded, that these fluctuations within 
the magnetic field could be driven by the residual numerical $\divB$ errors within the simulations, which do not make use
of the Dedner cleaning scheme.

\bFIGs
  \includegraphics[width=0.45\textwidth]{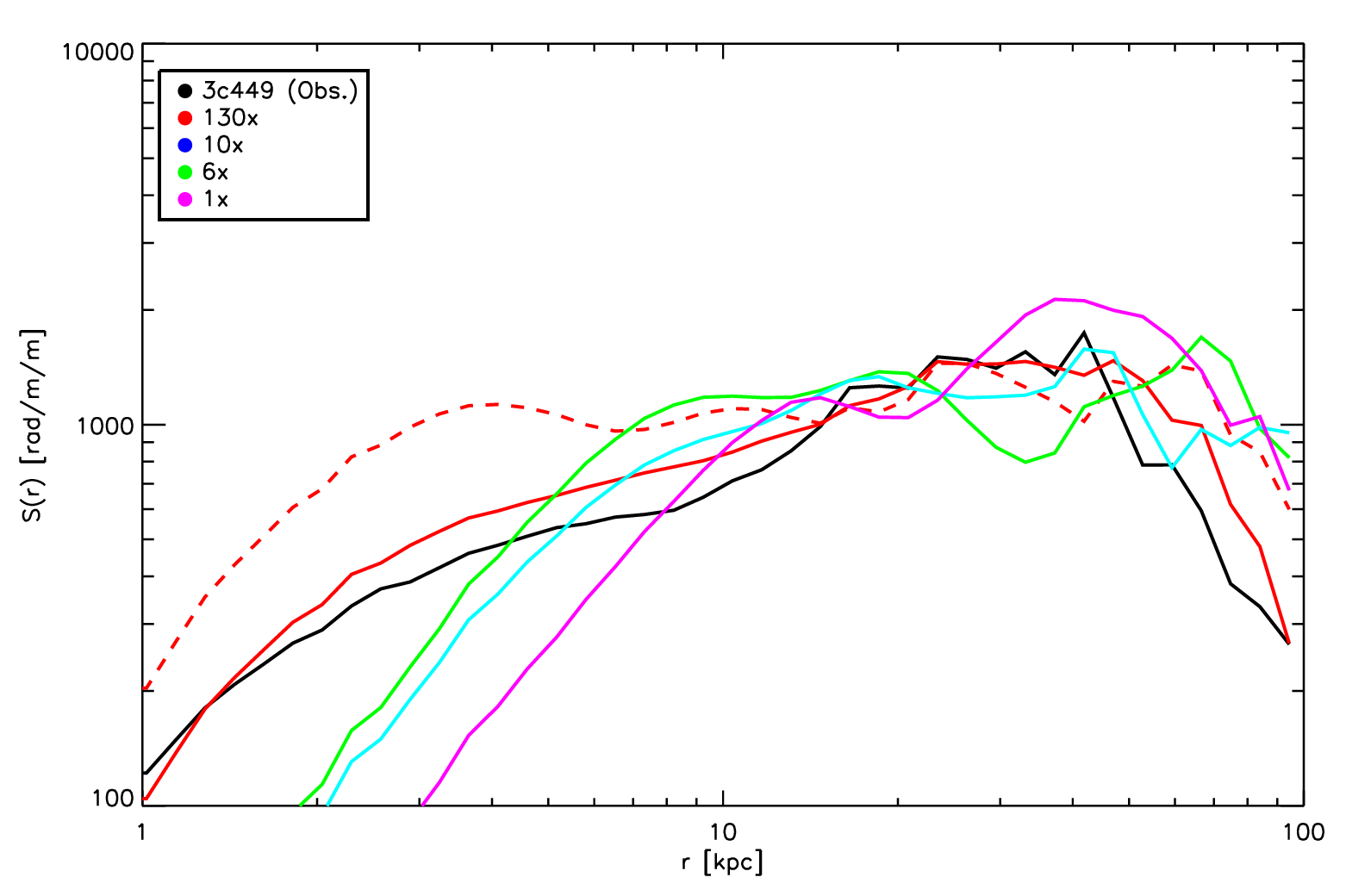}
\caption[Structure function for different resolution.]
{
Shown are structure functions $S$ calculated from the observed rotation measurement (black line) of 3C449 in \cite{1995A&A...302..680F} 
compared with the simulations at different resolutions with the cleaning scheme.
The structure functions are calculated from the map which has been cut to the shape of the region of the observed rotation measurements 
(as shown in \FIGp{rm_comp}).
The dashed blue line shows the 130x resolution run, without any cleaning scheme. 
}
\eFIGs{Str_cut_res}

Finally, \FIG{Str_cut_res}, shows the result where we calculated the structure function for the maps, cropped to the shape of the
observed rotation measurements (as shown in \FIGp{rm_comp}). The windowing due to the shape of the actual observations
leads to a decline of the correlation function at large scales, bringing the simulated structure functions in good agreement
with the observed one, whereas the small scales are not affected. Here, the simulations using the Dedner cleaning, at the highest 
resolution, show a remarkable match to the observations. However, the simulation without the Dedner cleaning scheme, shows a clear 
excess of small scale structures. The variability at small scales, indicate the need to include additional physics 
(i.e. ohmic dissipation, turbulent diffusion or viscosity) when further increasing the resolution. 
This will be needed to limit the turbulent cascade to not extend below the scales where the 
fluctuations are present in the observations.

\end{appendix}

\end{document}